\documentclass[runningheads]{llncs}

\usepackage{graphicx}
\usepackage[utf8]{inputenc}
\usepackage[english]{babel}
\usepackage{amsfonts}
\usepackage{amssymb}
\usepackage{amsmath}
\usepackage{tikz,pgfplots}
\pgfplotsset{compat=1.17}

\usetikzlibrary{arrows,decorations.markings}
\usetikzlibrary{arrows.meta}
\usepackage{algorithm}
\usepackage[noend]{algorithmic}
\usepackage{enumerate}

\begin{document}
\def\myDoubleLayout{0}

\title{Query Evaluation in DatalogMTL\\ -- Taming Infinite Query Results}
\author{
Luigi Bellomarini\inst{1}
\and Markus Nissl\inst{2}
\and Emanuel Sallinger\inst{2,3}
}
\authorrunning{L. Bellomarini et al.}
\institute{Banca d'Italia, Italy \\
\and TU Wien, Austria \\
\and University of Oxford, UK
}

\maketitle

\newcommand{\diamondplus}{%
  \raisebox{-.15ex}{\begin{tikzpicture}
    \useasboundingbox (-0.8ex, -0.8ex) rectangle (0.8ex, 0.8ex);
    \node (w) at (-0.8ex,0) {};
    \node (e) at (+0.8ex,0) {};
    \node (s) at (0,-.8ex) {};
    \node (n) at (0,+.8ex) {};
    \draw [-] (n.center) -- (e.center) -- (s.center) -- (w.center) -- (n.center);
    \draw [-] (n.center) -- (s.center);
    \draw [-] (e.center) -- (w.center);
  \end{tikzpicture}}}
\newcommand{\diamondminus}{%
  \raisebox{-.15ex}{\begin{tikzpicture}
    \useasboundingbox (-0.8ex, -0.8ex) rectangle (0.8ex, 0.8ex);
    \node (w) at (-0.8ex,0) {};
    \node (e) at (+0.8ex,0) {};
    \node (s) at (0,-.8ex) {};
    \node (n) at (0,+.8ex) {};
    \draw [-] (n.center) -- (e.center) -- (s.center) -- (w.center) -- (n.center);
    \draw [-] (e.center) -- (w.center);
  \end{tikzpicture}}}    
  
\newcommand{\Since}{\mathbin{\mathcal{S}}}
\newcommand{\Until}{\mathbin{\mathcal{U}}}
\newcommand{\DMTLFP}{DatalogMTL\textsuperscript{FP}}

\begin{abstract}
In this paper, we investigate finite representations of DatalogMTL models. 
First, we discuss sufficient conditions for detecting programs that have finite models.
Then, we study infinite models that eventually become constant and introduce sufficient criteria for programs that allow for such representation. 
We proceed by considering infinite models that are eventually periodic and show that such representation encompasses all \DMTLFP\ programs, a widely discussed fragment.
Finally, we provide a novel algorithm for reasoning over such finite representable programs.
\end{abstract}

\section{Introduction}
Query evaluation over temporal information is a fundamental problem in computer science, as temporal information occurs in many areas (e.g, the Internet of Things, financial systems, etc.). 
The Datalog-based language  DatalogMTL~\cite{DBLP:journals/jair/BrandtKRXZ18} has recently received significant attention in the research community as a powerful language for reasoning over temporal information.

DatalogMTL extends Datalog with operators from metric temporal logic (MTL) interpreted over the rational timeline and allows one to formulate expressions such as $\boxminus_{[0,5m]}\, \textit{Signal}(x)$ to state that the signal $x$ occurred continuously over the last 5 minutes or $\diamondminus_{[0,5m]}\, \textit{Signal}(x)$ to state that the signal $x$ occurred at least once in the last 5 minutes.
Recent work identified tractable fragments (for consistency checking) by using either $\boxminus$ or $\diamondminus$ operators in linear rules~\cite{DBLP:conf/ijcai/WalegaGKK20}.

\smallskip \noindent
\textbf{Challenge}.
However, such a framework allows for the formulation of rules that do not terminate and produce infinite models. 
\begin{example}
\label{ex:monday}
Consider the recursive rule $$\diamondminus_{[7d,7d]}\, \textit{Monday} \to \textit{Monday}$$ which shifts the validity of $\textit{Monday}$ by one week at each application. This clearly results in an infinite result, namely the infinite set of all future Mondays, and thus a non-terminating computation producing this result. At the same time, this \textit{is} a rule that represents natural domain knowledge, namely, that Mondays repeat every seven days.
\end{example}

\noindent
Producing infinite models is a frequent problem in many areas of Knowledge Representation and Reasoning, with countless solutions over the years. The core of such solutions typically lies in building \textbf{finite representations} of such infinite models---and subsequently using these finite models for solving the various tasks at hand.
If this were possible, we could solve tasks such as building a finite representation once and solving queries for arbitrary time points with little computational effort, unrolling the finite representation on demand to cover an arbitrary time period, etc.

\smallskip
\noindent
\textbf{Solution}.
In this paper, we investigate finite representations for DatalogMTL. This is in principle a wide area, which we divide into three types of finite representations:

\begin{enumerate}[i]
    \item \textit{finite models} (i.e., after a certain time point no facts hold)
    \item \textit{infinite models} that eventually become \textit{constant} (i.e., after a certain time point the set of facts remains the same)
    \item \textit{infinite models} that are eventually \textit{periodic} (i.e., after a certain time point there are repeating cycles of facts)
\end{enumerate}

\noindent
There is a persuasive mathematical analog of the above three types, namely, in number representations: for type (i) we have numbers such as 1.234, for (ii) e.g.,\ $1.2\dot{3}$, representing the number $1.233333...$ infinitely repeating, and for (iii) $1.\overline{234}$, meaning the number $1.234234...$ infinitely repeating.

\smallskip
\noindent
Our main contributions are:

\begin{itemize}
    \item We develop a \textbf{sufficient criterion to identify programs that have finite models, i.e.,  models of type (i)}. We call such programs \textit{harmless} and provide an algorithm for recognizing such harmless programs. We also show that for non-harmless programs, it is (non-trivially) database-dependent whether an infinite model is produced.
    \item For infinite models that eventually become constant, i.e., of type (ii), we see an interesting distinction between programs that contain $\boxminus$ and $\diamondminus$. For $\diamondminus$ we develop \textbf{a sufficient criterion called \textit{temporal linear} that allows finite representations of type (ii)}. We show that this criterion is, in some sense, optimal, as non-temporal linear programs may produce not-eventually constant models.
    \item For the case of $\boxminus$, we argue that \textbf{non-repetition of rule heads in Datalog programs is a sufficient criterion for type (ii)} finite representations. Slightly paraphrased, rules may not perform union operations. Finally, we show that when allowing both $\diamondminus$ and $\boxminus$, even \textit{both} criteria, i.e., temporal linear and union free, are not sufficient to guarantee type (ii) finite representations.
    \item This raises the question of what we can guarantee for programs that contain both $\diamondminus$ and $\boxminus$. We show that for \DMTLFP, the forward propagating fragment of DatalogMTL which exactly contains the temporal operators $\diamondminus$ and $\boxminus$, we can always guarantee \textbf{type (iii) finite representations}, i.e., repeating cycles of facts.
    \item We provide a \textbf{single algorithm} that can produce type (i), (ii), and (iii) finite representations for all the above restrictions, i.e., will take harmless programs and produce finite models; take temporal linear programs with $\diamondminus$ and produce type (ii) representations; etc. 
\end{itemize}

\smallskip\noindent
\textbf{Organization}. 
The remainder of this work is structured as follows: 
In Section~\ref{sec:preliminaries}, we introduce DatalogMTL. We discuss finite models for DatalogMTL in Section~\ref{sec:toolkit} and discuss infinite models in Section~\ref{sec:finite_fragments}. We introduce a new reasoning algorithm in Section~\ref{sec:algorithm}. We discuss related work in Section~\ref{sec:related_work} and finally, conclude the work in Section~\ref{sec:conclusion}.

\section{Preliminaries}
\label{sec:preliminaries}
In this section, we introduce the basic definitions of DatalogMTL under continuous semantics~\cite{DBLP:journals/jair/BrandtKRXZ18,DBLP:conf/ijcai/WalegaGKK19,DBLP:conf/ijcai/WalegaGKK20}.

\smallskip\noindent
\textbf{Syntax of DatalogMTL.}
Let \textbf{C} and \textbf{V} be disjoint sets of \emph{constants}, and \emph{variables}, respectively. A \emph{term} is either a constant or a variable. An \emph{atom} is an expression of the form $P(\boldsymbol{\tau})$, where $P$ is a predicate of arity $n \geq 0$ and $\boldsymbol{\tau}$ is a $n$-tuple of terms. A \emph{rule} is an expression of the form
\begin{equation*}
    A_1 \land \dots \land A_k \to B \hspace{2cm} \textrm{for } k \geq 0
\end{equation*}
where all $A_i$ and $B$ are literals that follow the grammar\footnote{We disallow in the head for undecidability reasons $\diamondminus$, $\diamondplus$, $\Since$ and $\Until$, and to ensure satisfiability $\bot$, as we focus on fact entailment}:
\if\myDoubleLayout1
\begin{align*}
    A ::= & \top \mid \bot \mid P(\boldsymbol{\tau}) \mid \boxminus_\varrho A \mid \boxplus_\varrho A \mid \\
    &  \diamondminus_\varrho A \mid \diamondplus_\varrho A \mid A \Since_\varrho A \mid A \Until_\varrho A \\
    B ::= & \top \mid P(\boldsymbol{\tau}) \mid \boxminus_\varrho A \mid \boxplus_\varrho A
\end{align*}
\else
\begin{align*}
    A ::= & \top \mid \bot \mid P(\boldsymbol{\tau}) \mid \boxminus_\varrho A \mid \boxplus_\varrho A \mid 
    \diamondminus_\varrho A \mid \diamondplus_\varrho A \mid A \Since_\varrho A \mid A \Until_\varrho A \\
    B ::= & \top \mid P(\boldsymbol{\tau}) \mid \boxminus_\varrho A \mid \boxplus_\varrho A
\end{align*}
\fi
where $\varrho$ is a \emph{non-negative} interval where an interval is represented in binary and of the form $\langle t_1,t_2 \rangle$, $t_1,t_2 \in \mathbb{Q}_2 \cup \{-\infty,\infty\}$, $t_1 \leq t_2$, and $\langle$ either open `$($' or closed `$[$' and $\rangle$ either open `$)$' or closed `$]$'. 
In the case of $\infty$ the corresponding end of the interval is defined as open.
The conjunction of $A_i$ is the rule \emph{body}, whereas $B$ is the rule \emph{head}. Instead of $\land$ we sometimes use ``$,$'' to denote conjunction. A predicate $p \in P$ is intensional (IDB), if $p$ occurs in the head of the rule whose body is not empty, otherwise $p$ is extensional (EDB). We define the left endpoint of the interval by $\varrho^{-}$ and the right endpoint by $\varrho^{+}$. The \emph{length of an interval} is defined as $|\varrho|=\infty$ if either $\varrho^{-}=-\infty$ or $\varrho^{+}=\infty$, otherwise $|\varrho| = \varrho^{+} - \varrho^{-}$.
A \emph{program} $\Pi$ is a finite set of rules and every program can be converted to \textit{temporal normal form} which contains only rules of the form:
\begin{align}
    P_1(\boldsymbol{\tau}_1) \land \dots \land  P_n(\boldsymbol{\tau}_n) \to & P_0(\boldsymbol{\tau}_0) & \hspace{0.5cm} (n \geq 0) \label{eq:horn_rule} \\
    P_1(\boldsymbol{\tau}_1) ~\mathcal{S}_\varrho~ P_2(\boldsymbol{\tau}_2) \to & P_0(\boldsymbol{\tau}_0) & \label{eq:since_rule} \\
    P_1(\boldsymbol{\tau}_1) ~\mathcal{U}_\varrho~ P_2(\boldsymbol{\tau}_2) \to & P_0(\boldsymbol{\tau}_0) & \label{eq:until_rule} \\
    \boxminus_{\varrho} P_1(\boldsymbol{\tau}_1) \to & P_0(\boldsymbol{\tau}_0) & \label{eq:box_rule} \\
    \boxplus_{\varrho} P_1(\boldsymbol{\tau}_1) \to & P_0(\boldsymbol{\tau}_0) & \label{eq:box_plus_rule}
\end{align}
In case we restrict $P_1$ in Rule (\ref{eq:since_rule}) and Rule (\ref{eq:until_rule}) to $\top$, we use instead the following rules in the normal form:
\begin{align}
     \diamondminus_{\varrho} P_1(\boldsymbol{\tau}_1)  \to & P_0(\boldsymbol{\tau}_0) & \label{eq:diamond_rule} \\
    \diamondplus_{\varrho} P_1(\boldsymbol{\tau}_1)  \to & P_0(\boldsymbol{\tau}_0) & \label{eq:diamond_plus_rule}
\end{align}
We call a program forward propagating, or in short \DMTLFP~\cite{DBLP:conf/aaai/WalegaKG19}, if it contains only rules of the form (\ref{eq:horn_rule}), (\ref{eq:box_rule}), and (\ref{eq:diamond_rule}).
A program $\Pi$ is \emph{bounded} if it does not contain $\top$ and all its intervals are bounded and \emph{union-free}, if each atom occurs at most once in the head of all rules. A program, rule or atom is ground, if it mentions no variables. Let $\mathrm{ground}(\Pi,D)$ be the grounding of the program with reference to only the constants in $\Pi$ and $D$.  A fact is represented by $A@\varrho$, where $A$ is a ground atom and $\varrho$ is the time-interval where $A$ holds. The dependency graph of $\Pi$ is defined as a direct graph $G(V,E)$, where $\Pi$ is in normal form, $v_p \in V$ for each predicate $P$ in $\Pi$ and $(v_p, v_q) \in E$, if there is a rule $r$ in $\Pi$ where $P$ is in the body and $Q$ in the head of the rule. We label an edge as special, if $r$ is of type (\ref{eq:since_rule})-(\ref{eq:diamond_plus_rule}). A program is \emph{recursive}, if the dependency graph is cyclic, and \emph {temporal recursive} if the cycle contains a special edge. Two predicates are \emph{mutually recursive} if they appear in the same cycle and \emph{temporal mutually recursive} if this cycle contains a special edge. A rule is \emph{temporal linear} if there is at most one body predicate that is mutually temporal recursive with the head in the dependency graph. An \emph{interval-labeling} of the graph, assigns to each edge the interval $[0,0]$ + $\hat{\varrho}$, where $\hat{\varrho} = \varrho$ for $\diamondminus_\varrho$, $\boxminus_\varrho$ and $\Since_\varrho$ and $\hat{\varrho} = -\varrho$ for $\diamondplus_\varrho$, $\boxplus_\varrho$ and $\Until_\varrho$, where $-\varrho = \{-t \mid t \in \varrho \}$ and $\varrho + \varrho' = \{ t + t' \mid t \in \varrho \land t' \in \varrho' \}$. A cycle is \emph{temporal-acyclic}, if the interval weight of the cycle is $[0,0]$, where the interval weight is defined as the sum of all interval-labels of the path. The interval weight $[t_1, t_2]$ is positive, if $0 \leq t_1 < t_2$.

\smallskip\noindent
\textbf{Semantics of DatalogMTL.}
The semantics are defined through an interpretation $\mathfrak{M}$. The interpretation specifies for each time point $t \in \mathbb{Q}$ and for each ground atom $P(\boldsymbol{a})$ whether $P(\boldsymbol{a})$ is true at $t$, in which case we write $\mathfrak{M},t \models P(a)$. We define inductively~\cite{DBLP:journals/jair/BrandtKRXZ18}:
\footnotesize
\if\myDoubleLayout1
\begin{align*}
\begin{aligned}
    & \mathfrak{M}, t \models \top & \textrm{ for each } t\\
    & \mathfrak{M}, t \models \bot &\textrm{ for no } t\\
    & \mathfrak{M}, t \models \boxminus_\varrho A & \textrm{iff } \mathfrak{M}, s \models A \textrm{ for all } s \textrm{ with } t-s \in \varrho \\
    & \mathfrak{M}, t \models \boxplus_\varrho A & \textrm{iff } \mathfrak{M}, s \models A \textrm{ for all } s \textrm{ with } s-t \in \varrho \\
    & \mathfrak{M}, t \models A~\Since_\varrho~A' & \textrm{iff } \mathfrak{M}, s \models A' \textrm{ for some } s \textrm{ with } t-s \in \varrho  \\ && \land~\mathfrak{M}, r \models A \textrm{ for all } r \in (s,t) \\
    & \mathfrak{M}, t \models A~\Until_\varrho~A' & \textrm{iff } \mathfrak{M}, s \models A' \textrm{ for some } s \textrm{ with } s-t \in \varrho \\ && \land~\mathfrak{M}, r \models A \textrm{ for all } r \in (t,s)  \\
    & \mathfrak{M}, t \models \diamondminus_\varrho A & \textrm{iff } \mathfrak{M}, s \models A \textrm{ for some } s \textrm{ with } t-s \in \varrho \\
    & \mathfrak{M}, t \models \diamondplus_\varrho A & \textrm{iff } \mathfrak{M}, s \models A \textrm{ for some } s \textrm{ with } s-t \in \varrho
\end{aligned}
\end{align*}
\else
\begin{align*}
\begin{aligned}
    & \mathfrak{M}, t \models \top & \textrm{ for each } t\\
    & \mathfrak{M}, t \models \bot &\textrm{ for no } t\\
    & \mathfrak{M}, t \models \boxminus_\varrho A & \textrm{iff } \mathfrak{M}, s \models A \textrm{ for all } s \textrm{ with } t-s \in \varrho \\
    & \mathfrak{M}, t \models \boxplus_\varrho A & \textrm{iff } \mathfrak{M}, s \models A \textrm{ for all } s \textrm{ with } s-t \in \varrho \\
    & \mathfrak{M}, t \models A~\Since_\varrho~A' & \textrm{iff } \mathfrak{M}, s \models A' \textrm{ for some } s \textrm{ with } t-s \in \varrho  \land~\mathfrak{M}, r \models A \textrm{ for all } r \in (s,t) \\
    & \mathfrak{M}, t \models A~\Until_\varrho~A' & \textrm{iff } \mathfrak{M}, s \models A' \textrm{ for some } s \textrm{ with } s-t \in \varrho  \land~\mathfrak{M}, r \models A \textrm{ for all } r \in (t,s)  \\
    & \mathfrak{M}, t \models \diamondminus_\varrho A & \textrm{iff } \mathfrak{M}, s \models A \textrm{ for some } s \textrm{ with } t-s \in \varrho \\
    & \mathfrak{M}, t \models \diamondplus_\varrho A & \textrm{iff } \mathfrak{M}, s \models A \textrm{ for some } s \textrm{ with } s-t \in \varrho
\end{aligned}
\end{align*}
\fi
\normalsize
Interpretation $\mathfrak{M}$ satisfies a fact $A@\varrho$, written $\mathfrak{M} \models A@\varrho$, if $M \models A@t$ for all $t \in \varrho$; a ground rule whenever it satisfies all atoms of the body it also satisfies the head of the rule;  a rule when it satisfies all groundings of the rule. An interpretation $\mathfrak{M}$ is a \emph{model} of a program if it satisfies all rules, and of a set of facts (which we call a database $D$) if it satisfies each of these facts. A program $\Pi$ and a database $D$ \emph{entail} a fact $A@\sigma$, written $(\Pi, D) \models A@\sigma$, if $A@\sigma$ for each model $\mathfrak{M}$ of $\Pi$ and $D$, and a set of facts $D'$, written $(\Pi, D) \models D'$, if each model of both $\Pi$ and $D$ is also a model of $D'$. We denote the set of facts $D'$ as $\Pi(D)$, if $D'$ is the intersection of all possible models of both $\Pi$ and $D$ (i.e., it is the minimum model). 

\section{Finite Models}
\label{sec:toolkit}
In this section, we classify the rules of a DatalogMTL program to distinguish between rules that terminate and rules that in general do not terminate when applying the immediate consequence operator~\cite{DBLP:conf/kr/WalegaZG21}.

Let us first point to the work of Walega et al.~\cite{DBLP:conf/kr/WalegaZG21}, who studied finite models of bounded DatalogMTL programs. From there, we know that non-recursive and temporal-acyclic bounded DatalogMTL programs always produce finite models. We build on this and derive a new sufficient condition that encompasses, among others, both previously introduced conditions, which we then use for classifying the rules of a DatalogMTL program.

\newcommand{\propInfiniteAnswersRule}{
Given a (bounded) DatalogMTL program $\Pi$ in temporal normal form, and a non-empty database $D$, it is database-dependent whether a temporal-cyclic program has a finite model.
}
\begin{proposition}
\label{thm:infinite_answers_rule}
\propInfiniteAnswersRule
\end{proposition}
\noindent
One core issue why termination is database-dependent for a lot of recursive programs containing temporal operators, is  due to the duality of the box operator (i.e., it behaves like a diamond operator in case the fact of the body is valid for a certain interval). Hence, a statement about termination (and reaching a finite model) by just considering the program without data is not possible in general. This is of course also true for \DMTLFP, as we still use the box operator.
\begin{corollary}
Given a (bounded) \DMTLFP program $\Pi$ in temporal normal form, and a non-empty database $D$, it is database-dependent whether a temporal-cyclic program has a finite model.
\end{corollary}

\noindent
Based on this observation, the goal is to find a sufficient condition for detecting a large group of programs that produce a finite model independently of the database. 
For this, we classify the rules into three groups: harmless rules, harmful rules, and dangerous rules. 

\begin{algorithm}[t]
\caption{Rule Classification Algorithm}
\label{alg:classification_rule_type}
\textbf{Input}: Program $\Pi$ \\
\textbf{Output}: Program $\Pi$, each rule assigned its type

\begin{algorithmic}[1]
\STATE $\mathit{DG} \gets \mathrm{dependecyGraph}(\Pi)$

\WHILE{$\mathit{DG}$ has been updated}
    \FOR{$\mathit{node}$ in $\mathit{DG}$}
        \IF{$\mathit{node}$ is finite according to Definition~\ref{def:finite_models}}
            \STATE mark $\mathit{node}$ as \textit{finite}
        \ENDIF
    \ENDFOR
\ENDWHILE
\FOR{$\mathit{rule} \in \Pi$} 
    \IF{Corresponding node of an atom in $\mathit{rule}$ is finite}
        \STATE mark $\mathit{rule}$ as \textit{harmless}
    \ELSIF{$\mathit{rule}$ is of type (\ref{eq:horn_rule})}
        \STATE mark $\mathit{rule}$ as \textit{harmful}
    \ELSE
        \STATE mark $\mathit{rule}$ as \textit{dangerous}
    \ENDIF
\ENDFOR
\end{algorithmic}
\end{algorithm}

\begin{definition}
\label{def:finite_models}
Let $\Pi$ be a bounded DatalogMTL program, then an edge in the dependency graph is marked as finite, if the corresponding rule contains a fact in the body who's corresponding node in the dependency graph is marked as finite and a node in the dependency graph is marked as finite, if (i) it is has no incoming edge (i.e., it is EDB), (ii) each incoming edge is marked as finite, (iii) after removing all finite edges and nodes in the dependency graph, the node is part of a strongly connected component, where each cycle is temporal-acyclic, (iv) the node is only part of simple cycles where each incoming edge to the cycle is an intersection with the cycle (assuming an empty database for the nodes in the cycle).
A rule is \emph{harmless} if at least one corresponding node of a body atom is finite.
A rule is \emph{harmful} if it is not harmless and of form (\ref{eq:horn_rule}) and \emph{dangerous} if it is neither harmless nor harmful.
A bounded DatalogMTL program $\Pi$ is \emph{harmless} if it contains only harmless rules.
\end{definition}

\noindent
The intuition behind this classification is as follows: Harmless rules have no impact on termination of the program and a program containing only harmless rules will result in a finite model. Harmful rules are also not critical if used in cycles, as they do not derive intervals at new time points, however in combination with dangerous rules they may not terminate. Dangerous rules form cycles that may lead to non-termination.

\newcommand{\theoremFiniteModels}{
Given a harmless bounded DatalogMTL program $\Pi$, the program terminates over an arbitrary database $D$ and produces a finite model.
}
\begin{theorem}
\label{thm:finite_models}
\theoremFiniteModels
\end{theorem}

\begin{figure}
    \centering
    \begin{tikzpicture}
            \node[shape=circle,draw=green] (A0) at (-1.5,1) {$X$};
            \node[shape=circle,draw=blue] (A1) at (-1.5,-1) {$Y$};
            \node[shape=circle,draw=red] (A) at (0,0) {$A$};
            \node[shape=circle,draw=red] (B) at (1.5,1) {$B$};
            \node[shape=circle,draw=red] (C) at (1.5,-1) {$C$};
            \node[shape=circle,draw=black] (D) at (3.5,1) {$D$};
            \node[shape=circle,draw=orange] (E) at (3.5,-1) {$E$};
            
            \path [->] (A0) edge node[midway, below, sloped] {$\mathit{{[1,2]}}$} (A1);
            
            \path [->] (A1) edge node[midway, above, sloped] {$\mathit{{[3,5]}}$} (A);
            
            \path [->] (A) edge node[midway, above, sloped] {$\mathit{{[0,0]}}$} (B);
            
            \path [->] (B) edge node[midway, below, sloped] {$\mathit{{[0,0]}}$} (C);
            
            \path [->] (C) edge node[midway, below, sloped] {$\mathit{{[0,0]}}$} (A);
            
            \path [->] (C) edge node[midway, above, sloped] {$\mathit{{[0,0]}}$} (D);
            
            \path [->] (D) edge[loop right] node[midway, right] {$\mathit{{[1,2]}}$} (D);
            
            \path [->] (D) edge node[midway, below, sloped] {$\mathit{{[0,0]}}$} (E);
            
            \draw [->] (E) to[out=0,in=90,looseness=4] node[midway, right] {$\mathit{{[1,2]}}$} (E);

    \end{tikzpicture}
    \caption{Identification of finite nodes, where green is type (i), blue (ii), red (iii), orange  (iv) and black is not finite (Definition~\ref{def:finite_models}).}
    \label{fig:finite_model_types}
\end{figure}
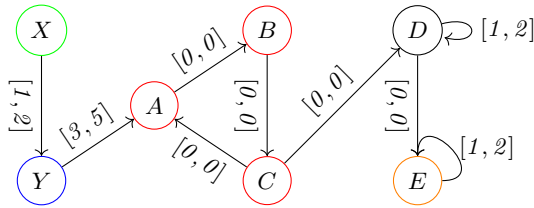

\smallskip\noindent
We use this theorem to suggest a classification algorithm (Algorithm~\ref{alg:classification_rule_type}). This algorithm takes as input a bounded DatalogMTL program $\Pi$ in normal-form and follows Definition~\ref{def:finite_models} by applying the defined cases to mark nodes and edges as finite. In Figure~\ref{fig:finite_model_types} we provide an example that shows the different cases. $X$ is finite due to type (i), $Y$ due to type (ii), $A$, $B$, and $C$ due to type (iii) as the cycle is temporal-acyclic, $D$ is not finite, and $E$ is finite due to type (iv).

We want to point out that a rule is harmless, if at least one corresponding node of a body atom is finite. This observation may be used in future work for further optimization, if one is interested in the entailment of a certain node, where one can limit the number of facts produced by the dangerous components. The main focus of this work is to find finite representations of such dangerous components to enable the termination of programs in general by detecting certain patterns:

\begin{definition}
\label{def:repetition_pattern}
Let $\Pi$ be a DatalogMTL program, $D$ a database, $P$ an arbitrary predicate in $\Pi$, and $T \in \mathbb{Q}$ a time point. 
A \emph{forward propagating repetition pattern} of $P$ is of size $n, n \in \mathbb{Q}^{>0}$, when $P(\tau)@t \in \Pi(D)$ iff $P(\tau)@(t+n*x) \in \Pi(D)$, where $T \leq t < T+n$, for all $P(\tau) \in \Pi(D)$ for all $x \in \mathbb{N}$ and all tuples $\tau$.
\end{definition}
\noindent
Intuitively, this means that the intervals of all valid facts after a time point $T$ are repeated linearly (cyclic). 
We say that a repetition pattern is \emph{empty} (i.e., there is a finite model), in case there is a time point $T$, such that for any predicate $P \in \Pi$, all tuples $\tau$ and for all $t \geq T$,  $P(\tau)@t \not \in \Pi(D)$. We will use this definition in the following sections.

\section{Finite Representations for Infinite Models}
\label{sec:finite_fragments}
In this section, we look at various fragments of DatalogMTL that can be used to investigate termination in detail. In particular, we focus on this and the following section on \DMTLFP\ and its sub-fragments, which have been studied intensively~\cite{DBLP:conf/aaai/WalegaKG19,DBLP:conf/kr/WalegaGKK20} as they allow us to derive facts in the future based on earlier facts. Note that the results also hold for the backward propagating fragment as they are symmetrical.

\subsection{Constant Interval Patterns}
\label{sec:finite_fragments_bounded_interval}

\noindent 
The studied fragments of this section are of high theoretical interest as they show a border between programs that produce -- independently of the database -- only finite models (previous section) and programs that are eventually periodic (next subsection). 
They are also representative of real-world phenomena that grow in duration until they eventually become constant. We give a (simplified) example here:

\begin{example}
\label{ex:earthovershootday}
The Earth Overshoot period is the part of the year during which humanity has consumed more of Earth's resources than it can supply. In a highly simplified formalization, this period extends by roughly one day per year (as it did in the years 2010-2019):
$$\diamondminus_{[365d,366d]}\, \mathit{OvershootPeriod} \to \mathit{OvershootPeriod}$$
It is easy to see that this produces an eventually constant, i.e., type (ii) representation.
\end{example}

\smallskip\noindent
We start by considering DatalogMTL\textsubscript{\diamondminus}, the fragment only allowing rules of type (\ref{eq:horn_rule}) and (\ref{eq:diamond_rule}).

\newcommand{\theoremDiamondMinus}{
Let $D$ be a database and $\Pi$ be a \emph{temporal linear} DatalogMTL$_{\diamondminus}$ program where each cycle in the dependency graph of $\Pi$ is temporal-acyclic or has a positive interval weight.
There exists a time point $T \in \mathbb{Q}$ s.t.\ for every $t > T$, $\alpha@t \in \Pi(D)$ iff $\alpha@T \in \Pi(D)$.
}
\begin{theorem}
\label{thm:diamond_minus}
\theoremDiamondMinus
\end{theorem}

\noindent
The construction of the theorem also provides some insights into the number of rule applications for a single cycle:
\begin{corollary}
Let $D$ be a database and $\Pi$ be a DatalogMTL program that consists of a single rule of the form $\diamondminus_{<t_{1},t_{2}>} P \to P$. This rule needs at most $n= \lfloor t_1/(t_2-t_1) +1 \rfloor$ applications, to reach a time point $T$ such that either for all $t' \geq T$ $P@t' \in \Pi(D)$ or for all $t' \geq T$ $P@t' \not \in \Pi(D)$.
\end{corollary}

\noindent
The theorem uses the two restrictions (i.e., temporal linearity and either temporal-acyclic or positive interval weight cycles) to limit the locations where the temporal operator $\diamondminus_{\langle t_1, t_2 \rangle}$ is allowed to prevent copying of the interval into the future (i.e., patterns resulting in $t_1=t_2$ with $t_1>0$, e.g. cases such as Example~\ref{ex:monday}). The following proposition shows that such patterns can be created when we remove the temporal linearity restriction, even if we forbid to use $t_1=t_2$.

\newcommand{\propDiamondMinusNonLinear}{
There exists a database $D$ and a DatalogMTL program $\Pi$ that only uses temporal operators $\diamondminus_{\langle t_1, t_2 \rangle}$ with $t_2 > t_1$ such that there exists \emph{no} time point $T \in \mathbb{Q}$ s.t.\ for every $t > T$, $\alpha@t \in \Pi(D)$ iff $\alpha@T \in \Pi(D)$. 
}
\begin{proposition}
\label{thm:diamond_minus_prop_non_linear}
\propDiamondMinusNonLinear
\end{proposition}

\noindent
As we will see in the corollary of the following results, the restrictions can be weakened to a certain extent by considering the interplay of the different intervals. Nevertheless, the chosen fragment is in a sense maximal as removing (and not replacing) either of the chosen restrictions (i.e., the linearity one or $t_2 > t_1$) yields finite representations of type (iii), which we discuss in the following section.

\smallskip\noindent
We continue to study DatalogMTL$_\boxminus$, the fragment only allowing rules of type (\ref{eq:horn_rule}) and (\ref{eq:box_rule}). As we have already discussed in Proposition~\ref{thm:infinite_answers_rule}, due to the duality of the box operator, the operator may behave like the diamond operator. In detail, this is the case for $\boxminus_{\langle t_1, t_2 \rangle}$ and a fact $P@\langle i1,i2 \rangle$, if $i1+t_2 < i2$, where we can replace $\boxminus_{\langle t_1, t_2 \rangle}$ with $\diamondminus_{[0, t_1 \rangle}$ (without considering the interval boundaries). This allows us to create infinite models by using the $\boxminus$-operator and hence again a restriction is necessary to produce type (ii) representations.

\newcommand{\propBoxMinus}{
Let $D$ be a database and $\Pi$ be a \emph{union-free} DatalogMTL$_\boxminus$ program, where each cycle in the dependency graph of $\Pi$ is temporal-acyclic or has a positive interval weight.
There exists a time point $T \in \mathbb{Q}$ s.t.\ for every $t > T$, $\alpha@t \in \Pi(D)$ iff $\alpha@T \in \Pi(D)$.
}

\begin{theorem}
\label{thm:box_minus}
\propBoxMinus
\end{theorem}

\noindent
Due to the duality of the box operator, we are also able to derive that each DatalogMTL\textsubscript{\diamondminus} program, where $t_1$ is restricted to $0$, results in a type (ii) representation (i.e., we do not require temporal linearity).

\begin{corollary}
\label{thm:box_minus_diamond}
Let $D$ be a database and $\Pi$ be a DatalogMTL program that only uses temporal operators $\diamondminus_{[0, t_2\rangle}$. There exists a time point $T \in \mathbb{Q}$ s.t.\ for every $t > T$, $\alpha@t \in \Pi(D)$ iff $\alpha@T \in \Pi(D)$.
\end{corollary}

\noindent
The theorem uses again two restrictions (this time union free instead of temporal linearity) to limit the locations of the temporal operators where $t1=t2$ is allowed to avoid the same issue as for the diamond operator. The new restriction forbids using the same rule head in multiple rules, as this allows to copy the intervals to adjacent places, which are merged, allowing to create a type (iii) representation. This is still possible if we disallow $t1=t2$ at all as the following proposition shows. 

\newcommand{\boxMinusCopyPattern}{
There exists a database $D$ and a \emph{temporal linear} DatalogMTL program $\Pi$ that only uses temporal operators $\boxminus_{\langle t_1, t_2 \rangle}$ with $t_2 > t_1$ such that there exists \emph{no} time point $T \in \mathbb{Q}$ s.t.\ for every $t > T$, $\alpha@t \in \Pi(D)$ iff $\alpha@T \in \Pi(D)$. 
}
\begin{proposition}
\label{thm:box_minus_copy_pattern}
\boxMinusCopyPattern
\end{proposition}

\smallskip\noindent
In this section, we studied infinite models that become constant for $\boxminus$ and $\diamondminus$. The propositions have highlighted the border, where such a model is and is not possible. We conclude this section by showing that the combination of both criteria (union free and temporal linear) will not result in type (ii) representations for $\diamondminus$ and $\boxminus$\footnote{We want to note that there exist other criteria, e.g., the combination of Theorem~\ref{thm:box_minus} and Corollary~\ref{thm:box_minus_diamond}.}.

\newcommand{\propBoxDiamond}{
There exists a database $D$ and a \emph{temporal linear, union-free} DatalogMTL program $\Pi$ that only uses temporal operators $\diamondminus_{\langle t_1, t_2 \rangle}$ and $\boxminus_{\langle t_1, t_2 \rangle }$ each with $t_2 > t_1$, such that there exists \emph{no} time point $T \in \mathbb{Q}$ s.t.\ for every $t > T$, $\alpha@t \in \Pi(D)$ iff $\alpha@T \in \Pi(D)$. 
}
\begin{proposition}
\label{thm:prop_box_diamond}
\propBoxDiamond
\end{proposition}

\subsection{Periodic Interval Patterns}
\label{sec:finite_fragments_linear_patterns}
In the previous subsection, we have identified the limits of constant intervals and noticed that already restricted combinations of $\diamondminus$ and $\boxminus$ programs may produce periodic intervals.
In this section, we study the combination of both operators ($\diamondminus$ and $\boxminus$), i.e., \DMTLFP, and show our main result that each model of such a program can be expressed by a type (iii) representation. 
Note that type (iii) representations allow to provide intervals of facts by multiple linear repetition patterns according to Definition~\ref{def:repetition_pattern} and are a superset of type (ii) representations as each constant interval is representable by a repeating one which is valid over the whole pattern length.

To derive the main result, we first provide two lemmas towards the theorem. The first lemma shows that it is sufficient to study a single strongly connected component (SCC) for repetition patterns and the second lemma shows that there exists a repetition pattern for an SCC. The main theorem combines the two results to show that each \DMTLFP can be represented by type (iii) representations.

\newcommand{\lemmaConnectedComponents}{
Let $\Pi$ be a \DMTLFP\ program. If there exists a repetition pattern for each strongly connected component in the dependency graph of $\Pi$, then there exists a repetition pattern of $\Pi$.
}
\begin{lemma}
\label{thm:connected_components}
\lemmaConnectedComponents
\end{lemma}

\noindent
Before we continue with the second lemma, we define an \emph{interval-shift}-labeled dependency graph. 

\begin{definition}
Let $\Pi$ be a grounded \DMTLFP\ program. The corresponding dependency graph is interval-shift-labeled, if each edge in the dependency graph originated from rule (\ref{eq:diamond_rule}) is labeled by the value $t_1$, of (\ref{eq:box_rule}) is labeled by value $t_2$, and for any other edge the value of the label is $0$.
\end{definition}

\noindent
The main idea of this labelling is to extract the amount each temporal operator shifts the left endpoint of a fact's interval in the future, which is essential for calculating the length of the repetition pattern.

\newcommand{\lemmaRepSizeN}{
Let $\Pi$ be a grounded \DMTLFP\ program which dependency graph contains a single strongly connected component. There exists a repetition pattern $p=lcm(R)$, if $p > 0$, else $p=1$ for definiteness, for each node in the dependency graph, where $R$ is a set, which contains for each simple cycle in an interval-shift-labeled dependency graph the sum $s$ of all edges contributing to the cycle, if $s < \infty$.
}

\begin{lemma}
\label{thm:rep_size_n}
\lemmaRepSizeN
\end{lemma}

\smallskip\noindent
Note that the calculated pattern length may not be the repetition pattern of the minimum length, as there may exist smaller patterns (i.e., a divisor of the pattern), which depends on the dominance of simple cycles. With these lemmas, we can show the main results of this subsection.

\newcommand{\sccRepPattern}{
Let $\Pi$ be a \DMTLFP\ program.
Let $D$ be a database.
Then the following holds:
For each predicate $P$ in $\Pi$, there exists a time point $T$ and integer $n$, such that $P(\tau)@(T+n*x) \in \Pi(D)$ iff $P(\tau)@T \in \Pi(D)$ for all $x \in \mathbb{N}$ and all tuples $\tau$.
}

\begin{theorem}
\label{thm:sccRepPattern}
\sccRepPattern
\end{theorem}

\noindent
Note that $n$ is the pattern length and results by the $lcm$ of the pattern length of each SCC (Lemma~\ref{thm:rep_size_n}).

\section{Reasoning Algorithm}
\label{sec:algorithm}
Theorem~\ref{thm:sccRepPattern} of Section~\ref{sec:finite_fragments_bounded_interval} shows that each \DMTLFP\ model is representable by a finite pattern. In this section, we will explore this theorem by presenting a reasoning algorithm that terminates for all \DMTLFP\ programs.

\begin{algorithm}[t]
\caption{Generic reasoning algorithm for \DMTLFP}
\label{alg:reasoning_algo}
\textbf{Input}: A \DMTLFP\ program $\Pi$ and database $D$ \\
\textbf{Output}: $\mathit{derivedFacts}$ and $\mathit{derivedPatterns}$

\begin{algorithmic}[1]

\STATE $\mathit{facts} \gets D$ and $\mathit{patterns} \gets \mathit{empty}$
\STATE Set $\mathit{pLength}$ to the pattern length (Theorem~\ref{thm:sccRepPattern})
\STATE $n \gets \lceil \mathtt{maxTimePoint}(D)/\mathit{pLength} \rceil$
\STATE $\mathit{nMin} \gets \lfloor \mathtt{minTimePoint}(D)/\mathit{pLength} \lfloor$
\STATE $\mathit{ruleGroups} \gets \mathtt{groupAndSortProgram(\Pi)}$.
\FOR{$\mathit{ruleGroup}$ in $\mathit{ruleGroups}$}
    \STATE $\mathit{prevNormFacts} \gets []$ and $\mathit{nPrev} \gets \mathit{nMin}$
    \LOOP
        \STATE Exhaustively apply \DMTLFP\ rules limited to intervals in the range $[\mathit{nPrev}*\mathit{pLength},(n+1)*\mathit{pLength})$ to $\{ \mathtt{extend}(patterns) \cup \mathit{facts} \}$
        \STATE $\mathit{normFacts} \gets \mathtt{normalize}(\mathit{facts},\mathit{pLength}, n)$
        \IF {$\mathit{normFacts}$ match $\mathit{prevNormFacts}$}
            \STATE $\mathit{patterns}.\mathit{addAll}(\mathtt{simplify}(\mathit{normFacts}))$
            \STATE \textbf{break}
        \ENDIF
        
        \STATE $\mathit{prevNormFacts} \gets \mathit{normFacts}$
        \STATE $\mathit{nPrev} \gets n$ and $n \gets n+1$
    \ENDLOOP
\ENDFOR
\end{algorithmic}
\end{algorithm}

Algorithm~\ref{alg:reasoning_algo} takes as input a \DMTLFP\ program $\Pi$ and a database $D$ and produces a set of facts $\mathit{facts}$ that contain facts of the form $P(\tau)@\varrho$ and a set of facts with its associated repetition information $\mathit{patterns}$ of the form $\{P(\tau)@\langle o_1, o_2 \rangle, \mathit{n}\}$, where the resulting intervals are given by $\langle o_1 + \mathit{x} * \mathit{pLength} , o_2 + \mathit{x} * \mathit{pLength} \rangle$ for all $x \in \mathbb{N}$, where $x \geq \mathit{n}$. First, we initialize the currently derived facts and patterns as well as calculate the pattern length according to the previous section (Lines 1-2). In Lines 3-4, we calculate the range where all database facts are located which have to be handled before any final pattern can be derived. Then (Line 5), we decompose the rules of the program into groups, such that each group exactly contains those rules which corresponding nodes in the dependency graph form a SCC and sort the groups in such a way, that if a group $X$ depends on $Y$ then $Y$ will be before $X$ in the list. 

In the following, we iterate over these groups and start by initializing in Line 7 the parameters for detecting patterns. We initialize the derived facts of the previous iteration to an empty list and the lower bound for the intervals of the derived facts to the start of the pattern that contains the start of the first interval of any entry in the database. 
The upper bound is given by Line 3 as the end of the pattern that follows the last interval of any entry in the database (but could be higher if previous $\mathit{ruleGroups}$ have already been computed). This ensures that all non-periodic facts of the database and of previous $\mathit{ruleGroups}$ are covered. 
We then exhaustively apply the \DMTLFP\ rules and derive all facts for the currently given interval range by the lower and upper bound of the pattern. The derivation is based on~\cite{DBLP:conf/aaai/WalegaKG19}, but due to space constraints and no added value, instead of repeating, we refer the reader to this work for details. The function $\mathtt{extend}$ unrolls the current patterns and computes the associated facts for the current interval range. Then, we check if a pattern is detected. For this, we $\textit{normalize}$ the derived facts by transforming each interval from $\langle t_1, t_2 \rangle$ to $\langle t_1-(n-1)*\mathit{pLength}, t_2-(n-1)*\mathit{pLength} \rangle$.
In case a pattern is detected, we stop the procedure and append the derived patterns to the output (Lines 11-13), otherwise we have to repeat the loop until a fixpoint is reached by deriving the facts for the intervals located in the next pattern (Lines 14-15). Note that at Line 12, we added a $\textit{simplify}$ function. The goal of this function is to detect intervals, which span the complete length and convert them into the form $[\mathit{pLength}*n,\infty)$.

Since this algorithm is based on Theorem~\ref{thm:sccRepPattern}, the known correctness of the derivation rules, and the consideration of every fact from the database for each $\mathit{ruleGroup}$, the following theorem establishes the correctness of our algorithm.

\newcommand{\algoReasoningSoundComplete}{
Algorithm~\ref{alg:reasoning_algo} outputs the answer to $(\Pi, D)$ for any database $D$ and any \DMTLFP\ program $\Pi$.
}

\begin{theorem}
\label{thm:algoReasoningSoundComplete}
\algoReasoningSoundComplete
\end{theorem}

\smallskip\noindent
\textbf{Optimizations.} In the presented algorithm we used for presentation purpose the same $\mathit{pLength}$ and $\mathit{n}$ for all $\mathit{ruleGroups}$. Note that these values can be computed per group, which will lead especially for the first groups to shorter pattern lengths as well as a possible lower value for $\mathit{n}$. In addition, after the detection of a pattern for a group, one can add an additional backtracking to check whether the pattern already existed for previous $n$ to reduce the $\mathit{n}$ also for the following components. The detailed discussion for the suggested and further optimizations is out of scope for this paper.

\begin{example}
Consider a database $D$ with a single fact $A@[0,1]$ and rules $\diamondminus_{[3,4]} A \to B$ and $\boxminus_{[3,4]} B \to A$. We retrieve $\mathit{pLength}=7$, $\mathit{nMin}=0$, $n=1$ and a single $\mathit{ruleGroup}$. We derive the following facts:
In iteration 1, we derive the facts $A@[0,1]$ and $B@[3,5]$.
In iteration 2, we derive the facts $A@[7,8]$ and $B@[10,12]$, which gets normalized to $A@[0,1]$ and $B@[3,5]$. The normalized facts match with the previous iteration and hence the patterns $A@[0+7n,1+7n]$ and $B@[3+7n,5+7n]$ for $n \geq 1$ are added and the loop terminates.
With optimizations, we would get $A@[0+7n,1+7n]$ and $B@[3+7n,5+7n]$ for $n \geq 0$.
\end{example}

\section{Related Work}
\label{sec:related_work}
DatalogMTL has been studied for the continuous, which also our work concentrates on, as well as for the pointwise~\cite{DBLP:conf/dlog/KikotRWZ18} semantics. Existing work primarily focused on the complexity results of model checking~\cite{DBLP:journals/jair/BrandtKRXZ18,DBLP:conf/ijcai/WalegaGKK19,DBLP:conf/aaai/CucalaWGK21,DBLP:conf/kr/WalegaCKG21}, including sub-fragments by restricting the operators~\cite{DBLP:conf/ijcai/WalegaGKK20} or by restricting the time points to the integer domain~\cite{DBLP:conf/kr/WalegaGKK20}. Walega et al.~\cite{DBLP:conf/aaai/WalegaKG19} also proposed an algorithm for fact entailment for deriving valid facts for a given set of time points for the forward propagating fragment, which builds the basis of our proposed algorithm. Recently, they~\cite{DBLP:conf/kr/WalegaZG21} studied fact entailment for finite materialisable programs by restricting DatalogMTL to bounded intervals and showed that there exists a finite model iff it is possible to construct a critical dataset and provided sufficient conditions for finite models. That work can be seen orthogonal to the presented work as we study those models that are not possible to be finitely materialized in that work. In addition, our work provides additional sufficient criteria for finite models. % Note: The work of Walega et al has been released at a conference after we submitted the first version of this paper on Arxiv.

The handling of infinite temporal models has received significant attention in the literature.
Temporal periodicity has been discussed in the context of rule-based systems by Chomicki and Imielinski~\cite{DBLP:conf/pods/ChomickiI88} who studied Datalog\textsubscript{1S}, a language that extends the predicates by one additional term to represent time and the successor function for this added term and identified there is a time point $T$ after which the derived facts repeat with a period of $p$. This language is a subset of DatalogMTL, where only punctual intervals (i.e., interval of the form $[t,t]$, and the temporal operator $\diamondminus_{[t,t]}$ is allowed (as only one temporal term is allowed). Tuzhilin and Clifford~\cite{DBLP:journals/is/TuzhilinC95} summarized different approaches of periodicity. They differ between strongly periodic sets, i.e., periods of repeated cycles (that is, periodicity as we discussed), and nearly periodic sets, i.e., periodicity that is happening or appearing at regular intervals (e.g., a meeting is once a week but not at a fixed time) and discussed possible integrations into TSQL2, but also restricted their focus to languages with point-based predicates. A general discussion about declarative rule-based, constraint-based, and symbolic approaches about temporal periodicity is provided by~\cite{DBLP:reference/db/Terenziani18a}. 

Temporal logic covers infinite models, among others, in the work around temporal tableaux~\cite{DBLP:conf/cav/Geilen03,DBLP:journals/logcom/Reynolds13,DBLP:conf/aiml/Reynolds14}, verification~\cite{DBLP:conf/fmcad/ClaessenS12,DBLP:conf/cav/CimattiGMRT19} or in general in the field of model checking~\cite{DBLP:conf/lics/OuaknineW05,DBLP:conf/formats/OuaknineW08,DBLP:conf/rv/HoOW14} where the focus is on
infinite reasoning in the sense of model checking, 
while our work studies the entailment of arbitrary facts where we derive new facts based on rules and existing facts. In addition, some of them focus on the pointwise semantics of MTL, while we focus on the continuous semantics, which provides a more natural understanding of temporal operators~\cite{DBLP:journals/acta/Reynolds16}. Recent worked studied the connection between finite and infinite traces~\cite{DBLP:conf/aaai/GiacomoMM14,DBLP:conf/ijcai/ArtaleMO19} of LTL and their differences in checking the satisfiability of formulas. Note that this problem is again different compared to ours.

\section{Conclusion}
\label{sec:conclusion}
In this work, we discussed finitely presentable fragments of DatalogMTL and introduced a reasoning algorithm that exploits the special properties of these fragments. In future work, we want to extend our work to full DatalogMTL. We also plan to study finite models more in detail
as they are a promising fragment for reasoning in knowledge graphs.

\section*{Acknowledgements}
The financial support by the Vienna Science and Technology Fund (WWTF) grant VRG18-013 is gratefully acknowledged.

\clearpage

\bibliographystyle{splncs04}
\bibliography{main}

\appendix
\section{Proofs}
In this section, we provide the proofs for the propositions, lemmas, and theorems introduced in the paper.

\subsection{Proofs for Section 3}

\medskip
\noindent
\textbf{Proposition \ref{thm:infinite_answers_rule}.}
\propInfiniteAnswersRule

\begin{proof}
It is possible to show that for an arbitrary temporal-cyclic DatalogMTL program $\Pi$ and a non-empty database $D$, it is only dependent on $\Pi$ whether it has a finite model does not hold. Here is an counterexample. Consider the following rule:
\begin{align*}
    \boxminus_{[3,7]} A \to & A
\end{align*}
In the first case, consider a database containing only a single fact $A@[0,1]$. Then we derive no additional fact, and the program terminates.
In the second case, consider a database containing only a single fact $A@[0,7]$. Then we derive $A@[7,10]$ (merged to $A@[0,10]$), $A@[7,13]$ (merged to $A@[0,13]$), \ldots, leading to the result $A@[0,\infty)$ after infinite applications. 
\if\myDoubleLayout1
\else
\qed
\fi
\end{proof}
The corollary follows immediately as the box operator is also part of the subfragment.

\medskip
\noindent
\textbf{Theorem \ref{thm:finite_models}.}
\theoremFiniteModels

\begin{proof}
We have to show that a Datalog program $\Pi$ in temporal normal form, where all rules are harmless, terminates.

We structure the proof as follows. First, we prove our first claim that for an arbitrary DatalogMTL program, a rule produces a finite model for the rule head, if at least one of the body atoms has a finite model. Then we focus on the individual claims for detecting finite nodes.  We show by case distinction that each case only produces a finite number of intervals and hence the program terminates and produces a finite model.

\textsc{Claim I}. We want to show for an arbitrary DatalogMTL program $\Pi$ and an arbitrary database $D$ that an arbitrary rule $r$ produces a finite model for the rule head, if at least one of the body atoms of $r$ has a finite model. 
Let $r:A_1,\ldots A_n \to B$ be an arbitrary rule from $\Pi$. Assume w.l.o.g. that $A_1$ has a finite model and let $\sigma_1$ be the set of intervals where $A_1$ holds and let $\sigma_b$ be the set of intervals where $B$ holds. By the semantics of the Horn rule, the intervals of $A_1, \ldots A_n$ are intersected and hence $\sigma_b \subseteq \sigma_1$. It remains to show that there exists a maximum number of intervals that can be added so that the procedure will terminate and we cannot produce a chain like $1, 1.5, 1.75, \ldots$ which does not terminate. By~\cite{DBLP:conf/ijcai/WalegaGKK19} we know that there exist only a finite (limited) number of possible intervals in a given range. This has been shown by constructing a ruler interpretation that constructs ruler intervals between and on time points, where the time points are limited by the intervals of the program and the database. Hence, we have shown that the rule produces only a finite model for the rule head.

\smallskip
\textsc{Claim II}. Let $\Pi$ be a bounded DatalogMTL program and $D$ an arbitrary database. We want to show that an arbitrary atom $A$ in $\Pi(D)$ has a finite model, if its corresponding node in the dependency graph of $\Pi$ has no incoming edges. We first show that $A$ is not derived by applying $\Pi$ on $D$. Assume $\Pi$ derives the value $A$. Then there must exist a rule $r \in \Pi$, such that $A_1,\ldots,A_n \to A$ for arbitrary $A_1,\ldots,A_n$. However, then there must exist an edge $(A_i,A)$ for all $1 \leq 1 \leq n$ in the dependency graph and hence $A$ has at least one incoming edge, which is a contradiction. Hence such a rule cannot exist and $A \in D$. As by definition all intervals have to be bounded in $D$, $A$ has a finite model.

\smallskip
\textsc{Claim III}.  Let $\Pi$ be a bounded DatalogMTL program and $D$ an arbitrary database. We want to show that an arbitrary atom $A$ in $\Pi(D)$ has a finite model, if all incoming edges of its corresponding node in the dependency graph are marked as finite. By definition, each incoming edge corresponds to a rule $r$ with a rule head $A$ which produces a finite model of $A$. The resulting model of $A$ is given by the union of all incoming models of the rules. As each rule produces a finite model, the union of these models is finite as well and hence also the atom $A$.

\smallskip
\textsc{Claim IV}. Let $\Pi$ be a bounded DatalogMTL program and $D$ an arbitrary database. We want to show that an arbitrary atom $A$ in $\Pi(D)$ has a finite model, if it is part of a SCC where each cycle is temporal-acyclic and all incoming edges are finite. 
As all incoming edges produce finite models, we can replace each edge by adding the finite model produced by the edge to the database $D$. This results in a single SCC, where we can apply the theorem of~\cite{DBLP:conf/kr/WalegaZG21}. Hence, the node in the SCC has a finite model, and hence $A$.

\smallskip
\textsc{Claim V}. Let $\Pi$ be a bounded DatalogMTL program and $D$ an arbitrary database. We want to show that an arbitrary atom $A$ in $\Pi(D)$ has a finite model, if the node is only part of simple cycles where each incoming edge to the cycle is an intersection with the cycle and $D$ is restricted to contain only facts for nodes not part of the cycle. We show this by contradiction. Assume the node is not finite. By definition, the node is part of a cycle and each node in the cycle has an incoming edge from the cycle. As each node has an incoming edge, no node of the cycle contains a fact from the database. So that at least one node in the cycle is valid at some time point, it requires an incoming edge from a different component of the dependency graph. Such edges are restricted to an intersection with the facts of a node inside the cycle. However, as shown, this node has no fact that holds at some time point and hence the intersection will not produce any valid fact as well. This results in a finite (empty) model, as no result is produced. Hence, we have a contradiction and the model is finite.

\smallskip
We have shown that all atoms produce a finite model where each corresponding nodes are marked as finite according to the definition. As \textsc{Claim 1} matches the definition of harmless rules, we have shown that a harmless bounded DatalogMTL program terminates and produces a finite model.
\if\myDoubleLayout1
\else
\qed
\fi
\end{proof}

\subsection{Proofs for Section 4.1}
\medskip
\noindent
\textbf{Theorem \ref{thm:diamond_minus}.}
\theoremDiamondMinus

\begin{proof}
We start this proof by showing that each DatalogMTL program that contains no temporal operator terminates. We then continue and show that a DatalogMTL program, which dependency graph has a single simple cycle which interval weight is not of the form $[t,t]$ with $t \neq 0$ has either a finite model or an infinite model that becomes constant. Then we restrict our proof to the diamond operator and show that the theorem holds for the restricted form of cycles. Finally, we show the theorem by also considering paths from and to cycles.

\smallskip
\textsc{Claim I}. An arbitrary DatalogMTL program $\Pi$ and an arbitrary database $D$ that does not contain a temporal operator terminates. In case the database is bounded (note the program is always bounded as it contains no temporal operators and hence no intervals), this follows from Theorem~\ref{thm:finite_models} as every cycle is temporal-acyclic. In case the database is unbounded, we have to show that we get an infinite model that is constant. First, we show that the model is not finite. Assume the model is finite. As the database is part of the model, the database has to be finite as well. This is a contradiction and hence the model is infinite. 
It remains to show that the infinite model is constant. This is, we have to show that the number of intersections and unions are bounded (which are the only allowed operations since only Horn rules are allowed). In short, as the database and the program contain only a finite number of facts/rules, only a finite number of intersections and unions can be produced. In detail, we define a slightly modified ruler that contains the time points $t_1,t_2$ for each fact $A@\langle t_1, t_2 \rangle$ of $D$ and a ruler-interval is the punctual interval for each time point (which is not $\infty$) on the ruler and $(i_1,i_2)$ for $i_1$ and $i_2$ be two consecutive time points on the ruler. As the database contains only a finite number of facts, the number of ruler-intervals is finite as well. By construction, if $A@t$, then $A@t'$, for $t'$ for every time point in the same ruler interval of $t$. This let us define the intersection of the two facts $A$ and $B$ as all ruler-intervals that are shared by $A$ and $B$ and the union as the ruler-intervals that are either in $A$ or $B$. As the number of ruler intervals is finite, and the number of possible atoms is finite (due to a finite number of facts in the database and rules in the program), we can derive a model, where at most for each atom $A$ and each ruler-interval $\langle t1, t2 \rangle$ the fact $A@\langle t_1, t_2 \rangle$ holds. Note that $t_1$ may be $-\infty$ and $t_2$ may be $\infty$. As the derivation of a finite number of intervals takes a finite amount of time, the program terminates.

\smallskip
\textsc{Claim II}. Let $G$ be the dependency graph of an arbitrary grounded DatalogMTL program $\Pi$ in normal form which dependency graph contains only a single simple cycle which interval-weight is not of the form $[t,t]$ with $t \neq 0$. We now show that this cycle either has a finite model or an infinite model that is constant. Note, that such a program, can only exist of linear rules, rules containing the box operator or rules containing the diamond operator. Since, Until and Horn rules with more than one body atom as well as Unions are not allowed, as this would result in more than one cycle. W.l.o.g., we replace each linear rule $P \to Q$ with $\diamondminus_{[0,0]} P \to Q$ (they are equivalent). As by the semantics, $\diamondminus_{[0,0]}$, $\boxminus_{[0,0]}, \diamondplus_{[0,0]}$, $\boxplus_{[0,0]}$ are equivalent, we insert w.l.o.g., additional auxiliary predicates to have in total $4n$ rules that create a cycle of the following form, where the predicate $P_{4n}$ is equal to the predicate $P_{0}$:
\begin{align*}
    \diamondminus_{\langle t_{4i+0,1}, t_{4i+0,2} \rangle} P_i \to P_{4i+1}\\
    \boxminus_{\langle t_{4i+1,1}, t_{4i+1,2} \rangle} P_i \to P_{4i+2}\\
    \diamondplus_{\langle t_{4i+2,1}, t_{4i+2,2} \rangle} P_i \to P_{4i+3}\\
    \boxplus_{\langle t_{4i+3,1}, t_{4i+3,2} \rangle} P_i \to P_{4i+4}
\end{align*}

Then by the semantics of the temporal operators, in case $P_0@\langle p_{0,1}, p_{0,2} \rangle$, we derive 
\begin{align*}
P_{4i+1}@\langle p_{4i+0,1} + t_{4i+0,1}, p_{4i+0,2} + t_{4i+0,2} \rangle\\
P_{4i+2}@\langle p_{4i+1,1} + t_{4i+1,2}, p_{4i+1,2} + t_{4i+1,1} \rangle\\ 
P_{4i+3}@\langle p_{4i+2,1} - t_{4i+2,2}, p_{4i+2,2} - t_{4i+2,1} \rangle\\ 
P_{4i+4}@\langle p_{4i+3,1} - t_{4i+3,1}, p_{4i+3,2} - t_{4i+3,2} \rangle     
\end{align*}
and hence the fact $P_0@\langle p_{0,1} + t_1, p_{0,2} + t_2\rangle$, where $t_1 = \Sigma_{i}^{n} t_{4i+0,1} + t_{4i+1,2} - t_{4i+2,2} - t_{4i+3,1}$ and $t_2 = \Sigma_{i}^{n} t_{4i+0,2} + t_{4i+1,1} - t_{4i+2,1} - t_{4i+3,2}$, if all intermediary intervals are valid (i.e., the left endpoint is lower or equal the right endpoint), otherwise the fact is not valid. By definition, $t_1 \neq t_2$, if $t1 \neq 0$. In case $t_1=t_2=0$, the cycle is temporal-acyclic and by the previous theorem we know that such a cycle has a finite model.

We now assume that our database consists of a single fact $P_0@\langle i_1, i_2 \rangle$. By applying the cycle $x$ times we derive the fact $P_0@\langle i_1 + (t_1*x), i_2 + (t_2*x) \rangle$, in case the interval is valid (the left endpoint is smaller or equal the right endpoint), otherwise this fact is not derived. We distinguish according to the possible value of $t_1$ and $t_2$ and assume that at least that there is one successful application (otherwise we have a finite model as no additional interval is produced for the fact). For a graphical representation of the cases, consider Figure~\ref{fig:case_distinction_proof_41_claim_2}. Note that we omitted in the list the case $t1=t2=0$ as in this case the program is temporal-acyclic, which we already covered.

\begin{itemize}
    \item \emph{$t_1 < t_2 \leq 0$}. As $t_1 < t_2$ the length of the interval is longer than the first interval (to be exact by a length $l=t_2-t_1$). Hence, there exists an $X$, such that after $X$ applications of the cycle, we will receive an interval $\langle i_1 + (t_1*X), i_2 + (t_2*X) \rangle$ that overlaps with the interval of the previous round $\langle i_1 + (t_1*(X-1)), i_2 + (t_2*(X-1)) \rangle$. In other words, the left endpoint of the interval of the previous iteration is part of the interval of the current iteration, which is given by $i_1 + (t_1) * (X-1) < i_2 + (t_1+l) * X$. By induction on the length, this overlapping also holds for all successive applications (i.e., for all $x \geq X$) and hence we can derive a fact $(-\infty, i_2 + t_2*X]$, an infinite model that is constant.
    \item \emph{$0 \leq t_1 < t_2$}. Similar to the previous case, but the right endpoint of the interval of the previous iteration is part of the interval of the current iteration, which is given by $i_2 + (t_1+l) * (X-1) > i_1 + t_1 * X$ and we derive a fact $[i_1 + t_1*y, \infty)$. Here, we also want to point out that we can calculate the number of applications $X$, until we derive the first overlapping which is given by $\lfloor (t_1-(i_2-i_1))/(t_2-t_1) +1 \rfloor$, where the numerator is the distance between the first two intervals and the denominator is the growth of the interval per application of the rule.\footnote{Note that this a more advanced form of the presented corollary of the paper considering not only a single rule, but a complete cycle. We omitted the extension to a single cycle due to lack of space and additional explanations required.}
    \item \emph{$t_1 < 0 < t_2$}. This combines both cases such that the previous interval is always a subinterval. Hence, we derive the fact $(-\infty, \infty)$.
    \item \emph{$t_2 < t_1 < 0$}. Normally in this case, the derived interval gets smaller, except it creates an overlapping interval with the existing one. This is the case if $i_1 \leq i_2+t_2$ and $i_1+t_1 \leq i_2$. In such a case, we can replace the interval with $\langle t_1,0]$ (as the left-endpoint is reduced by $t_1$, but the right endpoint stays within the interval) and apply the first case. In all other cases, we reach after $X$ applications a point such that $i_1 + (t_1*X) > i_2 + (t_2*y)$. This results in an invalid interval; no further fact is produced and hence results in a finite model.
    \item \emph{$t_2 < 0 = t_1$}. This case is special, as replacing with $\langle t_1,0]=[0,0]$ results in no additional interval, which results in a finite model.
    \item \emph{$t_2 < 0 < t_1$}. This case is special, as this results always in a subinterval of the previous interval and hence no additional interval is derived, which results in a finite model.
    \item \emph{$0 < t_2 < t_1$}. This case is similar to $t_2 < t_1 < 0$, except that we can replace the interval with $[0,t_2\rangle$ as the left endpoint stays within the interval in case an overlapping interval is derived.
    \item \emph{$t_2 = 0 < t_1$}. This case is special, as replacing with $[0,t_2\rangle=[0,0]$ results in no additional interval, which results in a finite model.
\end{itemize}

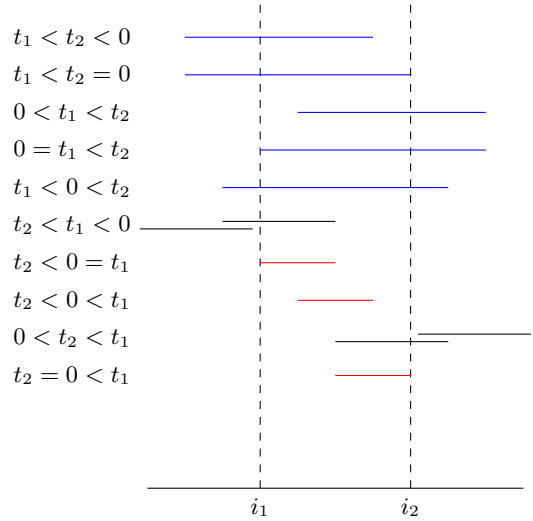
\begin{figure}[t]
    \centering
    
    \begin{tikzpicture}[scale=0.5]
    \draw (-5,-10)-- (5,-10);
    \draw[dashed] (-2,-10) -- (-2,3);
    \draw[dashed] (2,-10) -- (2,3);
    
    \draw (-7,-7) node {$t_2 = 0 < t_1$};
    \draw (-7,-6) node {$0 < t_2 < t_1$};
    \draw (-7,-5) node {$t_2 < 0 < t_1$};
    \draw (-7,-4) node {$t_2 < 0 = t_1$};
    \draw (-7,-3) node {$t_2 < t_1 < 0$};
    \draw (-7,-2) node {$t_1 < 0 < t_2$};
    \draw (-7,-1) node {$0 = t_1 < t_2$};
    \draw (-7,0) node {$0 < t_1 < t_2$};
    \draw (-7,1) node {$t_1 < t_2 = 0$};
    \draw (-7,2) node {$t_1 < t_2 < 0$};

    \draw[color=red] (0,-7)-- (2,-7);
    \draw (0,-6.1)-- (3,-6.1); 
    \draw (2.2,-5.9)-- (5.2,-5.9);
    \draw[color=red] (-1,-5)-- (1,-5);
    \draw[color=red] (-2,-4)-- (0,-4);
    \draw (-5.2,-3.1)-- (-2.2,-3.1);
    \draw (-3,-2.9)-- (0,-2.9);
    \draw[color=blue] (-3,-2)-- (3,-2);
    \draw[color=blue] (-2,-1)-- (4,-1);
    \draw[color=blue] (-1,0)-- (4,0);
    \draw[color=blue] (-4,1)-- (2,1);
    \draw[color=blue] (-4,2)-- (1,2);

    \draw (-2,-10.5) node {$i_1$};
    \draw (2,-10.5) node {$i_2$};
    
    \end{tikzpicture}
    
    \caption{Graphical representation of the different cases. The initial interval is between $i_1$ and $i_2$, the individual cases are described on each line. Blue lines show intervals that are increasing of size (which at some point overlap with the previous interval). Red lines show intervals that do not produce any new time point. Black intervals mark the dual ones, where we draw both cases, either an extending interval, or a shortened interval outside. }
    \label{fig:case_distinction_proof_41_claim_2}
\end{figure}

We know extend the reasoning to an arbitrary number of facts. Trivially, if the database contains no facts, no interval, and a finite model is derived and if the database contains multiple facts, there may be a coalescing of produced intervals from the cycles and existing intervals. This has no impact on the reasoning applied to a single interval, as either all intervals result in a point where no additional valid interval is produced or where multiple intervals are coalesced into a single interval where we can again start with the reasoning for a single interval. As there are only a finite number of starting facts, the number of intervals to consider does not increase and only decreases in the case of coalescing, we again derive either a finite or a constant infinite model. Hence, such cycles always lead either to a constant infinite or a finite model.\footnote{Note that we have shown a more complex version, allowing to combine different temporal operators in a simple cycle as long as the interval length of the cycle is zero or not punctual.}

\smallskip
\textsc{Claim III}. We show that for a database $D$ and a temporal linear DatalogMTL program $\Pi$ that uses only the temporal operator $\diamondminus_{\langle t1, t2 \rangle}$ and which dependency graph of $\Pi$ contains a single SCC either derives a finite model or an infinite model that is constant. Trivially, if it contains a SCC without a cycle, the program contains no rules and hence  we derive a finite model (if the database is bounded) or an infinite model that is constant (if the database is unbounded).
Now we focus on the case that the SCC has multiple nodes. We already have shown in \textsc{Claim II} that this is true for a simple cycle. In the case of this theorem, the only relevant case is $0 \leq t1 < t2$ and as no other temporal operators are allowed by the semantics of the temporal operator. In case the database is not empty, we know that this cycle results in the interval $[T,\infty)$. In case the database is empty, we have a finite model.

W.l.o.g., we assume that our first cycle of the dependency graph has a positive interval-weight. 
We show by induction that adding such a path to the cycle results either in a finite model or an infinite model that is constant. Assume we have added $n$ paths, and we add an additional path to the cycle. 

We first consider that this path intersects with the existing SCC. 
We observe the following: Due to the temporal linearity constraint, it is only possible to add an path that intersects with the existing SCC in the following way: The path creates an own cycle that contains no temporal operator and starts and ends at the same node in the existing SCC, or it adds a path between two edges of a previously added cycle that does not contain any temporal operators. It is not possible to add a path that connects two nodes of any cycle containing temporal operators, as this will break the temporal linearity constraint (as the intersection has multiple body atoms and this results in two body atoms being mutually temporal recursive). This is, the cycle that may produce an infinite model is bounded by the facts of the intersecting cycle added to the nodes. In order that an intersection results in a valid fact, both incoming edges require facts from the database. As the intersection cycle does not have any temporal operator, the intervals are bounded by the facts from the database. In case there is no fact $[T, \infty)$, then the resulting intersection is finite else it is either finite or infinite with a constant interval as the infinite fact is given from the database, and the intersection with the cycle either yields a point, where no additional valid interval is produced (as the intersection does not fire), or the intersection of the incoming intersection cycle is of type $[T, \infty)$ so that the normal reasoning in the simple cycle applies (following the previous claim). This is, the intersection propagates through each existing simple cycle recursively, which updates the cycle either by a bounded interval (coming from the intersection) or stays a constant infinite model.

It remains to study the union operation. We have to distinguish the following possibilities: (i) a path from the temporal cycle to an intersection cycle, (ii) a path in the temporal cycle, (iii) a path in the intersection cycle, and (iv) an edge from an intersection cycle to the temporal cycle, or (v) a path starting and ending at the same node in an intersection cycle and (vi) a path starting and ending at the same node in a union cycle. 

Case (i) is not allowed as this violates the temporal linearity condition as due to the intersection there are now two body atoms which are mutually recursive. 

In Case (ii), we split the temporal cycle into two subcycles, where we can apply the reasoning per cycle (the existing one) and the new one established by the added path according to \textsc{Claim II}. As both cycles have to be either temporal-acyclic or have a positive interval weight, the same reasoning applies per cycle. The only point we have to consider is that the intervals of both cycles are merged at the target node (i.e., this node where the new path ends). There we have to distinguish between different merge possibilities. In the case of the existing cycle having a constant infinite model, then the new path has no influence on the cycle and the nodes in the new cycle will also get a constant infinite model as there is no intersection in the newly added path. In the case that the existing cycle has a finite model, then there are different possibilities. The new cycle bypasses the limiting factor and creates a constant infinite model, which is propagated also to the existing cycle or this cycle also has a finite model. As the limitation is caused by an intersection, it usually stays a finite model, except if the intersections have infinite facts which have not overlapped with the existing cycles, but due to the merging of the intervals there now existing some pattern which causes a constant infinite model. Note that only a constant infinite model is possible, due to the positive interval required in each cycle.

Case (iii) bypasses some intersections in the intersection cycle. This may yield intervals, which have been filtered out in the cycle but no additional time points are added as each added path is not allowed to contain any temporal operator due to the linearity condition. Hence, the possible produced intervals are bounded to the database of the intersection cycle and the reasoning follows the discussed reasoning for intersection cycles.

In Case (iv), by the definition of an intersection cycle (and the other union cases), the only temporal incoming edge to such an cycle is the initial edge of a temporal cycle. Hence, such an edge just creates additional values, which we forward into the temporal cycle. By considering the graph, one notices that this is the same as case (ii), by constructing the graph in a different way. First, add the union edge (see Case ii), and then apply the intersection edge to this cycle (which we discussed for intersections).

In Case (v), the path is not allowed to contain any temporal operator as otherwise the temporal linearity constraint of the intersection cycle is broken. Hence, it consists only of linear rules and is not modifying any temporal intervals, leading to no change in the existing cycles.

In Case (vi), we can split the reasoning into two subcycles, where we can apply similar reasoning as in case (ii), except that the new path cannot have any intersection. Hence, in case there exists any database entry for a node in the newly added path or the starting node contains any fact, the starting node will result always in a constant infinite model due to the new cycle in case the path contains at least one temporal edge, otherwise the added path has no influence on any temporal interval.

This concludes the proof of \textsc{Claim III}, showing that each SCC of the chosen form either results in a finite or constant infinite model.

\smallskip
\textsc{Claim IV}. In this claim, we finally connect different SCCs and show that the theorem holds.
We have three different edge types we have to connect SCC: (i) rule of type (\ref{eq:diamond_rule}), (ii) rule of type (\ref{eq:horn_rule}) and (iii) unions. W.l.o.g., we can introduce auxiliary nodes, such that each rule has at most two body atoms, and we can study each rule on its own. Case (i) is not relevant for an incoming edge to a cyclic SCC as this can either be (ii) or (iii), but has to be considered for the outgoing edges of an SCC. For (i), it is trivially to see that each option produces only finite models or constant infinite models as the application of the diamond operator to an interval results in an unbounded interval only in case either the temporal operator or the fact has an unbounded interval. As there are only a finite number of facts, this operation also produces only a finite number of facts. Similar reasoning can be applied to (ii) and (iii), if they are not incoming edges to a cyclic SCC. 

It remains to show that the incoming edges to a cyclic SCC result either in a finite model or a constant infinite model. For (iii) it is trivial, as this has been shown by \textsc{Claim III} as we considered an arbitrary database, and we can simply extend the database with the facts resulting from this incoming edge. It remains to study (ii). This is again similar two \textsc{Claim III}, but here, we add to a node $p$ an intersection cycle $p \to q \to p$ for some fresh predicate $q$ and initialize $q$ with exactly the intervals from the previous SCC. This concludes the proof and shows that each model is either finite or infinite with constant intervals.

This is, in case it is finite, there is a point $T$ after which no fact holds, or if infinite there is a time point $T \in \mathbb{Q}$, such that $\alpha@t \in \Pi(D)$, iff $\alpha@T \in \Pi(D)$ for all $t > T$.
\if\myDoubleLayout1
\else
\qed
\fi
\end{proof}

\medskip
\noindent
\textbf{Proposition \ref{thm:diamond_minus_prop_non_linear}.}
\propDiamondMinusNonLinear

\begin{proof}
It is possible to show that for arbitrary datasets $D$ and an arbitrary DatalogMTL program $\Pi$ that only uses temporal operators $\diamondminus_{t_1, t_2}$ with $t_2 > t_1$ there exists \emph{a} time point $T \in \mathbb{Q}$ s.t.\ for every $t > T$, $\alpha@t \in \Pi(D)$ iff $\alpha@T \in \Pi(D)$ does not hold. Here is a counterexample: Consider a database containing the facts $A@[0,3]$ and $B@[2,4]$ and the following rules:
\begin{align*}
    A, B \to C \hspace{1cm} \diamondminus_{[3,5]} C \to A \hspace{1cm} \diamondminus_{[5,6]} C \to B
\end{align*}
This produces the following facts: $A@[5n,5n+3]$, $B@[5n+2,5n+4]$, and $C@[5n+2,5n+3]$, for all $n \in \mathbb{N}$.
\if\myDoubleLayout1
\else
\qed
\fi
\end{proof}

\medskip
\noindent
\textbf{Theorem \ref{thm:box_minus}.}
\propBoxMinus

\begin{proof}
The proof is similar to the proof of $\diamondminus$. In \textsc{Claim II}, the only case possible is of form $0 < t_2 < t_1$ by the semantics of the box operator (Note that we refer throughout the proof to $t_1$ and $t_2$ of the claim and not the temporal-operator to stay consistent with the claims we reuse.) . This case contains the duality of the box operator, given in case the cycle does not create a gap between the current and the next interval so that the interval length is not reduced and we can transform the interval to the form $[0, t_2 \rangle$. By \textsc{Claim IV} we also know that the combination of different SCCs has no influence on the theorem, as the same reasoning applies here. 
In this theorem it remains to prove the equivalence of \textsc{Claim III} for the restriction of this theorem for a cyclic SCC.

We now study the options when we add additional rules to the SCC by induction and assume that we already have a SCC of size $n$. We only have to study a single case, namely, adding a path with $\boxminus$-operator which results in an intersection (as unions are not allowed). This results in three options: (i) the SCC is already reaching a point, where no further interval is produced, then the new intersection only reduces the allowed intervals and the SCC may reach this point earlier, (ii) the SCC produces an interval of the form $[0, t_2 \rangle$, but due to the intersection, the new interval contains a gap and hence a finite model is reached or (iii) the added intersection still behaves like an interval of the form $[0, t_2 \rangle$. In order that (iii) is possible, we observe that there is a point $T'$ such that for every $t > T'$ each node in the SCC creates overlapping intervals. This is the case, as in case there is one node which creates overlapping intervals, then the following nodes have to create overlapping intervals as well, and hence the incoming edges of an intersection have overlapping intervals as well, which extends the intersection so that the resulting interval of the intersection is also overlapping.
Hence, there exists a time point $T \in \mathbb{Q}_{\ge 0}$ s.t.\ for every $t > T$,  $\alpha@t \in \Pi(D)$ iff $\alpha@T \in \Pi(D)$. Note that, if one fact in the SCC is not extending, the complete SCC is not extending and there exists a time point where no additional fact is produced.

Note for the Corollary: The only possibility for getting the overlapping intervals is the case, if the box operator behaves like a diamond operator which has an interval of the form $[0, t_2 \rangle$. This is the same as if we consider a DatalogMTL program that only uses the operator $\diamondminus_{[0,t2\rangle}$, where we get the same behavior. What remains open is why we dropped the union-free requirement. As the intervals are only extending, and every node depends on the incoming edges, in all cases the interval is limited to the smaller extension of the interval from the beginning of some fact (this is the main difference to the box operator, where we can create gaps, which we cannot in this case). This is, if there is no intersection that has an unextended interval (i.e., a length of $0$), the interval gets extended (and as stated above, this cannot be stopped as it is propagated along the SCC). In case there is such an intersection with a unextended interval, finite intervals may be produced, as in case no other extending interval is on the path, the intervals are limited to the database and its applicable rules. 
\if\myDoubleLayout1
\else
\qed
\fi
\end{proof}

\medskip
\noindent
\textbf{Proposition \ref{thm:box_minus_copy_pattern}.}
\propBoxMinus

\begin{proof}
It is possible to show that for an arbitrary dataset $D$ and an arbitrary \emph{temporal linear} DatalogMTL program $\Pi$ that only uses temporal operators $\boxminus_{t_1, t_2}$ with $t_2 > t_1$ there exists \emph{a} time point $T \in \mathbb{Q}$ s.t.\ for every $t > T$,  $\alpha@t \in \Pi(D)$ iff $\alpha@T \in \Pi(D)$ does not hold. Here is a counterexample:
Consider a database containing the facts $A@[2,5]$ and the following rules:
\begin{align*}
    A \to B \hspace{1cm} \boxminus_{[1,2]} A \to B \hspace{1cm} \boxminus_{[10,12]} B \to A
\end{align*}
This will produce the following facts: $A@[10n+2,10n+5]$, and $B@[10n+2,10n+6]$ for all $n \in \mathbb{N}$.
\if\myDoubleLayout1
\else
\qed
\fi
\end{proof}

\medskip
\noindent
\textbf{Proposition \ref{thm:prop_box_diamond}.}
\propBoxDiamond

\begin{proof}
It is possible to show that for an arbitrary dataset $D$ and an arbitrary temporal linear, union-free DatalogMTL program $\Pi$ that only uses temporal operators $\diamondminus_{t_1, t_2}$ with $t_2 > t_1$ and $\boxminus_{t_1, t_2}$ with $t_2 > t_1$, there exists \emph{a} time point $T \in \mathbb{Q}$ s.t.\ for every $t > T$, such that $\alpha@t \in \Pi(D)$ iff $\alpha@T \in \Pi(D)$ does not hold. Here is a counterexample: Consider a database containing the facts $A@[0,3]$ and the following rules:
\begin{align*}
    \diamondminus_{[5,6]} A \to B \hspace{2cm} \boxminus_{[4,5]} B \to A
\end{align*}
This will produce the following facts: $A@[10n,10n+3]$ and $B@[10n+5,10n+9]$ for all $n \in \mathbb{N}$.
\if\myDoubleLayout1
\else
\qed
\fi
\end{proof}

\subsection{Proofs for Section 4.2}

\medskip
\noindent
\textbf{Lemma \ref{thm:connected_components}.}
\lemmaConnectedComponents

\begin{figure}[t]
    \centering
    \includegraphics[width=0.9\columnwidth]{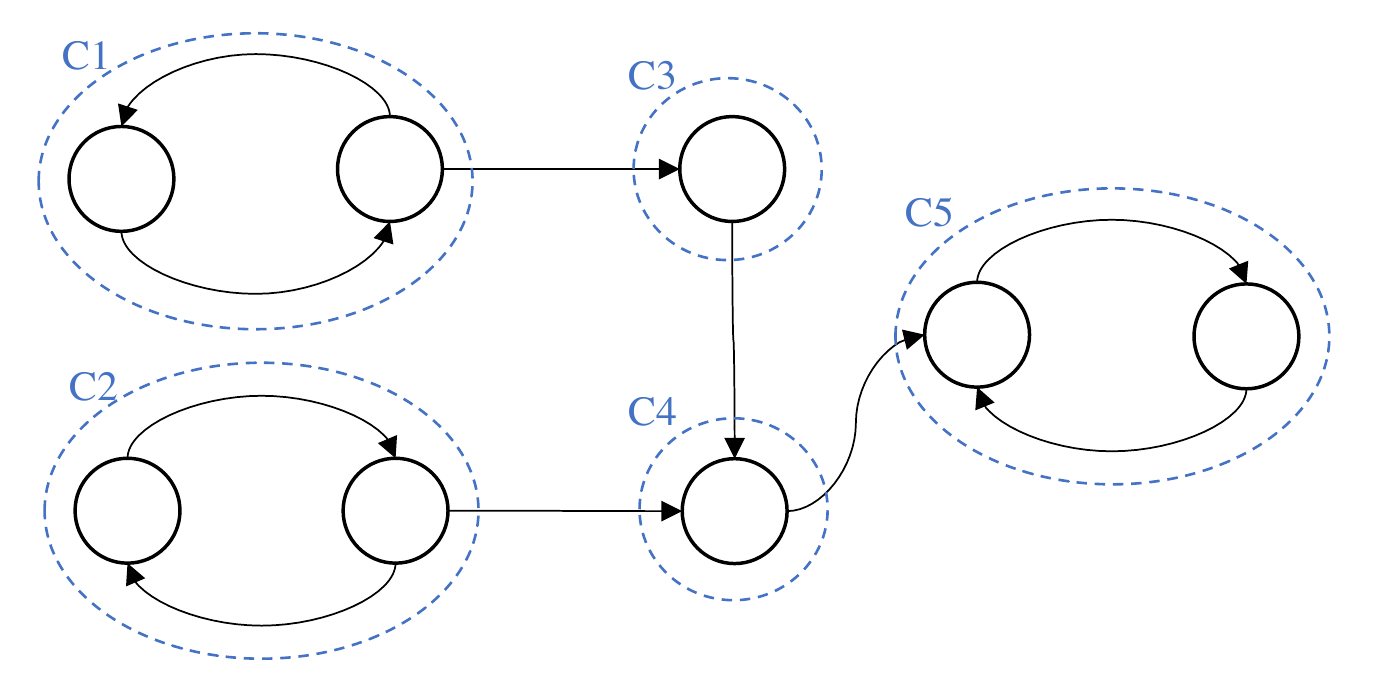}
    \caption{Induction Component Types: C1, C2 are type 1, C3 is type 2, C4 is type 3, and C5 is type 4}
    \label{fig:connected_components}
\end{figure}
\begin{figure}[t]
    \centering
    \begin{tikzpicture}[scale=0.4]

    \draw[->] (0,0)-- (18,0); 
    
    \foreach \x in {0,1,2,3,4,5,6,7,8,9,10,11} {
        \draw ({\x*1.5},0.5) -- ({\x*1.5},-0.5) node[below] {$\x$};
    }
    
    \node at (-1, 4)  (a) {A};
    \node at (-1, 3)  (b) {B};
    \node at (-1, 2)  (c) {C};
    \node at (-1, 1)  (d) {D};
        
    \foreach \x in {0,1,2,3,4,5} {
        \foreach \y in {4} {
            \draw ({(2*\x)*1.5},{\y}) -- ({((2*\x)+1)*1.5},{\y});
            \draw[fill=black] ({(\x*2)*1.5},{\y}) circle (0.25);
            \draw[fill=white] ({((\x*2)+1)*1.5},{\y}) circle (0.25);
        }
    }
    
    \foreach \x in {0,1,2,3} {
        \foreach \y in {3} {
            \draw ({(3*\x)*1.5},{\y}) -- ({((3*\x)+1)*1.5},{\y});
            \draw[fill=black] ({(\x*3)*1.5},{\y}) circle (0.25);
            \draw[fill=white] ({((\x*3)+1)*1.5},{\y}) circle (0.25);
        }
    }
    
    \foreach \x in {0,1} {
        \foreach \y in {2} {
            \draw ({(6*\x)*1.5},{\y}) -- ({((6*\x)+1)*1.5},{\y});
            \draw[fill=black] ({(\x*6)*1.5},{\y}) circle (0.25);
            \draw[fill=white] ({((\x*6)+1)*1.5},{\y}) circle (0.25);
        }
    }
    
    \foreach \x in {0,1} {
        \foreach \y in {1} {
            \draw ({(6*\x)*1.5},{\y}) -- ({((6*\x)+1)*1.5},{\y});
            \draw[fill=black] ({(\x*6)*1.5},{\y}) circle (0.25);
            \draw[fill=white] ({((\x*6)+1)*1.5},{\y}) circle (0.25);
        }
    }
    
    \foreach \x in {0,1} {
        \foreach \y in {1} {
            \draw ({((6*\x)+2)*1.5},{\y}) -- ({((6*\x)+5)*1.5},{\y});
            \draw[fill=black] ({((6*\x)+2)*1.5},{\y}) circle (0.25);
            \draw[fill=white] ({((6*\x)+5)*1.5},{\y}) circle (0.25);
        }
    }
    
    \end{tikzpicture}
    \caption{Intersection $A,B \to C$ and union $A \to D$, $B  \to D$}
    \label{fig:intersection_union_example}
\end{figure}
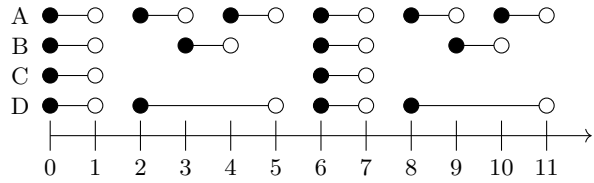
\begin{figure}[t]
    \centering
    \includegraphics[width=.9\columnwidth]{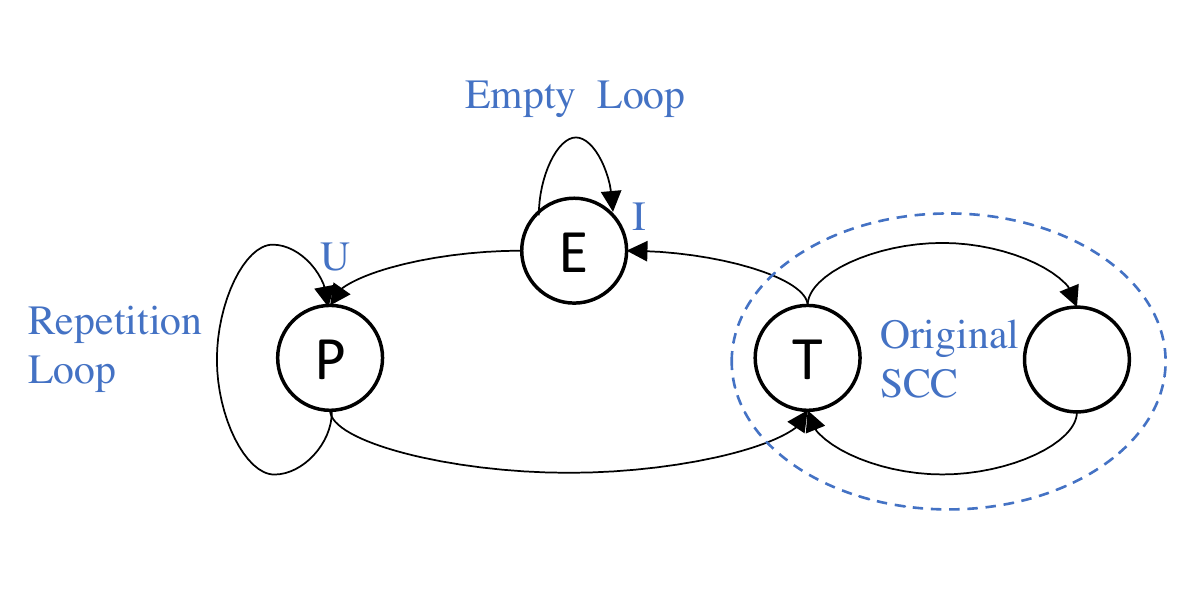}
    \caption{Simulation of a pattern input of a SCC inside a single SCC}
    \label{fig:simulation_scc_single}
\end{figure}
\begin{proof}
We prove by induction. Let us assume that we have shown the repetition pattern for $n$ strongly connected components (let's say $c_k$, $0 < k < n$) and we want to show that component $c_n$ has also a repetition pattern, where $c_n$ only depends on strongly connected components $c_k$. Then we have to distinguish between four cases (visualized in Figure~\ref{fig:connected_components}):
\begin{enumerate}
    \item The component has no incoming edges from a previous connected component. This can be reduced to the base case (a single connected component) and is not further discussed is this Lemma.
    \item $c_n$ is a component forming a cycle, i.e, each incoming edge is from a strongly connected component $c_{k}$ with a pattern length of $l_k$. This edge only appears together with an intersection or a union as input to the node in the cycle. In general, the intersection can be interpreted as a ``constant'' filtering operation that removes certain timestamps and can be seen as an upper bound of the produced atoms. In contrast, in case of a union, the rule head at least contains the provided repetition pattern and can be seen as a lower bound of produced values of this node. As the input can be simulated by a single SCC, it is sufficient to show for a single strongly connected component that there exists a repetition pattern. See Figure~\ref{fig:simulation_scc_single} for the construction of such a single SCC, where $T$ is the target node, i.e., the incoming node is from a strongly connected component $C_{k}$, $E$ is the node, which database is empty, $P$ is the node which database contains the values of the repetition pattern, $I$ marks an intersection and $U$ a union of the incoming edges. That is, the output of node $E$ is always empty, hence the input of node $P$ is always forced to be the repetition pattern, which is then also the input of node $T$.
    \item $c_n$ is a single node, and it has two incoming edges with weights $0$ (it is either a union or an intersection) and pattern lengths $l_1$ and $l_2$ from strongly connected component $c_{k_1}$ and $c_{k_2}$. This trivially has a repetition pattern, again. The new repetition length is determined as $l = lcm(l_1,l_2)$ and the incoming patterns are either intersected or merged and will start again together after $lcm(l_1,l_2)$ steps (see Figure~\ref{fig:intersection_union_example} for an example).
    \item $c_n$ is a single node, and it has one incoming edge with weight $w$ (given by the labeling of the temporal operator, i.e., $t_2$ for $\boxminus$, $t_1$ for $\diamondminus$) from a strongly connected component $c_k$ with a pattern length of $l$. This trivially has a repetition pattern. The pattern is just shifted by $w$ time units and the pattern is modified by the temporal operator. Note that the pattern may be final after one additional application as an extension will create consistently overlapping facts in the next pattern.
\end{enumerate}
This concludes the proof that each strongly connected component can be analyzed individually for the existence of repetition patterns.
\if\myDoubleLayout1
\else
\qed
\fi
\end{proof}

\medskip
\noindent
\textbf{Lemma \ref{thm:rep_size_n}.}
\lemmaRepSizeN

\begin{proof}
This proof is structured into two steps. In the first step, we show that a SCC has a node that allows to split the SCC into two SCCs. In the second step, we show by induction that the path length of each simple cycle is sufficient to compute the length of the repetition pattern.

\smallskip
\noindent
\textsc{Claim 1.} Each SCC with at most two incoming and two outgoing edges per node (which we can assume w.l.o.g.) with at least one node containing two outgoing edges has (1) a node $N$ that contains a critical node (i.e., removing this nodes split the SCC into two components), or (2) a node $N$ where both outgoing paths reach a node with two incoming edges not later than reaching a node with two outgoing edges. 

\smallskip
\noindent
A SCC either contains (1) or not, hence we only have to proof (2), when the SCC does not contain (1). We show this by contradiction.
First, we remove all nodes that contain only a single incoming and outgoing edge, as these nodes are not relevant and combine the two edges into a single edge (i.e., edges $(n_i, n_j)$, $(n_j,n_k)$ to $(n_i,n_k)$). Second, we assume that both outgoing edges reach a different node (otherwise this trivially holds, as the two edges are both incoming edges fulfilling the statement). 
Now, assume there exists no such node $N$. Then we have three options: (i) both edges end at a node with two incoming edges, which contradicts the assumption, (ii) one edge reaches a node with two incoming edges and one with two outgoing edges, or (iii) both edges reach a node with two outgoing edges. In case the node has two outgoing edges, the only incoming edge is the edge from the current considered node. By definition, each node has to reach all other nodes of the SCC, hence there must exist a node, where both outgoing edges have to add an edge to an existing node (which already has at least one incoming node) and now has two incoming edges, being of the form (i). This is, in case such a node would exist, one always has to add an additional node, which does not close the SCC, which results in a contradiction that we have an SCC. Hence, such a node must exist.

\noindent
\smallskip
In the following, we call such a node $N$ a split node.

\medskip
\noindent
\textsc{Claim 2.} Each SCC with at most two incoming and two outgoing edges per node with at least one node containing two outgoing edges can be reduced to a set of single simple cycles that contain all simple cycles of the original SCC.

\smallskip
\noindent
This follows by iteratively applying Claim 1. Case (1) naturally is a divider of simple cycles and splits the SCC into two components, where we retrieve per component all simple cycles in the following iterations. In case (2), we split exactly at a single split node (i.e, we remove the path from the split node until the first node contains two incoming edges). As simple cycles only pass a node once, all simple cycles are retained, some in one and some in the other SCC.

\smallskip
\noindent
In the following, we call the first nodes on the paths that contain two incoming edges after the split node merge nodes.

\medskip
\noindent
\textsc{Claim 3.} The repetition pattern of a single simple cycle is given by the sum $s$ of the edge labels, where each edge originated form rule (\ref{eq:diamond_rule}) is labeled by the value $t_1$, of (\ref{eq:box_rule}) is labeled by value $t_2$, and for any other edge the value of the label is 0.

\smallskip
\noindent
This follows from the first part of the proof of Theorem 4, where we sum up the shifting of the intervals. As for each round the interval always only either increases (up to the length $s$, then just copied), decreases (to an empty interval), or stays the same, we have a repetition pattern of length $s$.

\medskip
\noindent
\textsc{Claim 4.} The repetition pattern of a SCC is the least common multiple of all simple cycles.

\smallskip
\noindent
This can be shown by induction on the number of paths of a SCC.
By Claim 3, we have shown the base case and we assume in the following that we have shown it for SCC of $n$ paths. We want to show that this also holds for SCCs, where we add one additional path. 
By Claim 1, we know that each SCC has a split node, and by Claim 2 we know that each SCC can be reduced to SCCs containing simple cycles and such a splitting preserves all simple cycles. We use one iteration of Claim 2 to create two SCCs $X$ and $Y$ of $n$ paths with repetition lengths $x$ and $y$ (i.e., we exactly remove a single path for each SCC). Now it remains to show that combining the two SCCs leads to a pattern length of $lcm(x,y)$, which we show by case distinction.
\begin{itemize}
    \item \textbf{Merge Node = Split Node - Intersection}. In this case, we have two strongly connected components that are only connected via this node. This is, both cycles have the dominant role and once an interval is not valid, it will propagate this information every $x$ or $y$ steps. This is, it only reduces the number of possible created intervals in $X$ and $Y$. As both cycles start their repetition pattern together again after $lcm(x,y)$ and there is no change in the pattern anymore (i.e., no additional intervals are removed), the following iterations will always follow the same pattern. Hence, the new repetition pattern is $lcm(x,y)$.
    \item \textbf{Merge Node = Split Node - Union}. This case is similar to the previous case with the only difference that instead of removing intervals, additional intervals are added. Hence, the new repetition pattern is $lcm(x,y)$.
    \item \textbf{One Merge Node - Intersection} In such a case, both paths end at the same merge node. This is, the pattern of the merge node is propagated to the split node (in the same way for both patterns), then it is split at the split node, and then combined again at the merge node. This is, the merge node's pattern is propagated equally on both paths until the split node. This is, an intersection only removes intervals. As both patterns repeat together again after $lcm(x,y)$, one can apply similar reasoning as for the previous cases and hence the new repetition pattern is $lcm(x,y)$.
    \item \textbf{One Merge Node - Union} This case is similar to the previous case with the only difference that instead of removing intervals, additional intervals are added. Hence, the new repetition pattern is $lcm(x,y)$.
    \item \textbf{Two Merge Nodes - Intersection,Intersection} In this case, the two paths end at two different nodes, merge nodes $M1$ and $M2$. One can observe the following. Both nodes, $M1$ and $M2$ are part of $X$ and $Y$ and so there must be a path from $M1$ to $M2$ and a path from $M2$ to $M1$ without using the split node $S$, since by removing the path from $S$ to $M1$ requires that $M2$ reaches $M1$ and removing the path from $S$ to $M2$ requires that $M1$ reaches $M2$ without using $S$. This is the pattern is propagated equally along all paths except the paths between $S$ and $M1$ and $S$ and $M2$, i.e., the patterns are computed by the union of the paths and since all paths are the same, they only depend on the patterns of the merge nodes. As both nodes are intersection and hence only removes intervals and both patterns repeat together again after $lcm(x,y)$, one can apply similar reasoning as for the previous cases and hence the new repetition pattern is $lcm(x,y)$. 
    \item \textbf{Two Merge Nodes - Union,Union} This case is similar to the previous case with the only difference that instead of removing intervals, additional intervals are added. Hence, the new repetition pattern is $lcm(x,y)$.
    \item \textbf{Two Merge Nodes - Intersection,Union} This case is the more complex case as the intervals are filtered at the intersection node and added at the union node. Hence, one has to prove that the pattern is still $lcm(x,y)$ and not a multiple of it. The key observation is that there is a path between $M1$ and $M2$ in both directions. This is the intersection propagates the pattern to the union via the existing nodes and (optionally) via the split node and the union similarly propagates the pattern to the intersection node. In case there is a path from the split node via the merge node to the split node with no node containing two incoming edges (except for the merge node), such a path creates a cycle, which has no additional input. Hence, in case such a node is a union, it acts as a lower bound for the pattern and in case it is an intersection, it acts as an upper bound for the pattern. Hence again we can calculate the pattern as $lcm(x,y)$. This pattern can be even further extended to cycles which contain two incoming edges, in case the operation is the same as the merge node. 
    
    Now we consider the remaining cases and continue with a similar argument, where we inspect each node. For this, we mark the graph as follows:
    \begin{enumerate}
        \item We mark the union merge node as union dominating.
        \item We mark the intersection merge node as intersection dominating.
        \item We mark each node, where all incoming edges are union dominating, union dominating.
        \item We mark each node, where all incoming edges are intersection dominating, intersection dominating.
        \item We mark each union node, where at least one incoming edge is union dominating, union dominating.
        \item We mark each intersection node, where at least one incoming is intersection dominating, intersection dominating.
        \item We repeat 3-6.
    \end{enumerate}
    We observe that the split node is either union or intersection dominating. This is, as $M1$ and $M2$ are connected, each union or intersection along the path requires as input either both values from $M1$ or $M2$ or one value from $M1$ and $M2$. In the latter case, the merge type of these nodes decides the race for this node. This is, one can find a cycle, which is purely decided either on the union or the intersection merge node. In case it is union-dominating, the pattern is a lower bound, in case it is intersection-dominating, the pattern is an upper bound\footnote{Remind that the upper bound increases until it reaches a fixpoint per iteration and the lower bound decreases until it reaches a fixpoint}, hence we can calculate the pattern as $lcm(x,y)$.
\end{itemize}
This concludes our proof that there exists a pattern of length $lcm(x,y)$.
\end{proof}

\medskip
\noindent
\textbf{Theorem \ref{thm:sccRepPattern}.}
\sccRepPattern
\begin{proof}
This theorem follows from the previous lemmas. By Lemma~\ref{thm:rep_size_n}, we know that a repetition pattern exists for a single connected component and by Lemma~\ref{thm:connected_components}, we know that if a repetition pattern exists for a single connected component, then it exists for all connected components of a program. Together this proves the theorem. 
\if\myDoubleLayout1
\else
\qed
\fi
\end{proof}

\medskip
\noindent
\textbf{Theorem \ref{thm:algoReasoningSoundComplete}.} \algoReasoningSoundComplete

\smallskip\noindent
\begin{proof}
We have to show the soundness and completeness of the theorem.
First, we show the soundness. We only derive facts that are based on the derivation rules of \DMTLFP\ which mimic the semantics. The underlying execution is based on previous work~\cite{DBLP:conf/aaai/WalegaKG19} which showed the soundness of their algorithm, hence also this program is sound. For the patterns, we have proven in Section~\ref{sec:finite_fragments_linear_patterns} in the theorem and lemmas that the correct pattern length as well as that such a pattern exists as soon as it is repeated after the last fact of the database has been handled. This is, our patterns only contain valid repetition patterns.
We now have to show completeness. This means, we have to derive at least all valid facts up to some time point $T$ and after $T$ we have to cover at least all valid patterns. Again, the derived facts are a result of the underlying derivation rules and due to the fixpoint semantics of Datalog, we do not miss any case. It remains to show that every pattern is derived. We know (by Theorem and Lemma), that we reach a point, where all successive patterns derive the same facts (normalized to the pattern length). This is exactly what we do in the algorithm and hence we derive all patterns. The same condition also guarantees termination of the algorithm (as we break the loop in case two successive patterns match).
\end{proof}

\end{document}